\documentclass[reprint,onecolumn,aps,showpacs,nofootinbib,notitlepage]{revtex4-2}
\pdfoutput=1
\usepackage[normalem]{ulem}
\usepackage{graphicx}
\usepackage{psfrag}
\usepackage{caption}
\usepackage{subcaption}
\usepackage{float}
\usepackage{amsmath}
\usepackage{amssymb}
\usepackage{amsthm}
\usepackage{mathtools}
\usepackage{dsfont}
\usepackage{xcolor}
\usepackage{comment}
\usepackage{caption}
\usepackage{bigints}
\usepackage{geometry}
\usepackage{scalerel}[2016/12/29]

\usepackage{lineno}
\usepackage{ulem}

\newcommand{\be}{\begin{equation}}
\newcommand{\ee}{\end{equation}}
\newcommand{\ra}{\rightarrow}

\def\bc{\begin{center}}
\def\ec{\end{center}}
\def\bea{\begin{eqnarray}}
\def\eea{\end{eqnarray}}

\newcommand{\at}[2][]{#1|_{#2}}

\newcommand{\deletetext}[1]{\iffalse{{\color{red}{#1}}}\fi}
\newcommand{\newtext}[1]{{\color{black}{#1}}}

\begin{document} 

\title{Current fluctuations of a self-interacting diffusion on a ring}

\author{Francesco Coghi}
\email{francesco.coghi@su.se}
\affiliation{Nordita, KTH Royal Institute of Technology and Stockholm University, Hannes Alfvéns väg 12, SE-106 91 Stockholm, Sweden}

\date{\today}


\begin{abstract}

We investigate fluctuations in the average speed or current of a self-interacting diffusion (SID) on a ring, mimicking the non-Markovian behaviour of an agent influenced by its own path. We derive the SID's phase diagram, showing a delocalisation-localisation phase transition from self-repelling to self-attracting. Current fluctuations are analysed using: (i) an adiabatic approximation, where the system reaches its stationary distribution before developing current fluctuations, and (ii) an original extension of level 2.5 large deviations for Markov processes \newtext{combined with perturbation theory}. \newtext{Both methods} provide lower bounds to current fluctuations, \newtext{with the former tighter than the latter in all localised regimes, and both equally} tight in the self-repelling region. Both methods accurately estimate the asymptotic variance and suggests a phase transition at the onset of the localised regime.

\end{abstract}

\maketitle

\tableofcontents

\section{Introduction}

In the study of stochastic processes and statistical physics, the fluctuations of Markov (memory-less) processes are now well understood, with comprehensive theories like large deviation theory providing a robust framework for their description~\cite{DenHollander2000,Touchette2009,Dembo2010,Jack2020a}. However, many real-world systems exhibit long-range correlations in time, making them inherently non-Markovian. These systems require models that account for memory effects, which can significantly alter their dynamic behaviour. Non-Markov processes are not only better suited for understanding real-world phenomena where memory plays a crucial role, but they also arise naturally from the coarse-grained dynamics of Markov systems. In this work, we aim to advance the study of fluctuations in non-Markov processes by focusing on Self-Interacting Diffusions (SIDs), an exemplary class of processes characterised by long-range temporal correlations.

We consider an SID on the ring $[0,2\pi)$ described by the process $( \theta_t )_{t \geq 0}$  solution of the Stochastic Differential Equation (SDE)
\begin{equation}
\label{eq:SIDEmp}
d\theta_t = \sqrt{D} dW_t + dt \int_0^{2 \pi} 2 c \sin (\theta - \theta_t) \rho_t(\theta) \, d\theta \, ,
\end{equation}
where $D>0$ is the diffusion coefficient of the Brownian motion $W_t$ on the ring, $c \in \mathbb{R}^+$ a constant determining the magnitude of the periodic drift and $\rho_t$ the empirical occupation measure of the process, which is a scalar field marking the fraction of time spent by the process on each and every point of the ring, up to time $t$ (in Sec.\ \ref{sec:Model} we mathematically formalise this concept). Roughly, the process evolves stochastically on a ring and its position at time $t$ is affected by the whole path of the process up to time $t$ through $\rho_t$. We refer the reader to Sec.\ \ref{sec:Model} for further details on the process.   We are interested in the study of the average speed, or current, observable 
\begin{equation}
\label{eq:CurrObs}
\Omega_t = \frac{1}{t} \int_0^t d\theta_s \, ,
\end{equation}
and aim to characterise its fluctuation behaviour in the long-time limit using large deviation theory methods. 

SIDs have been introduced in the probability community in~\cite{Benaim2002} as examples of continuous-time path-interaction (or reinforcement) processes to model growing polymers. SIDs extend a class of self-interacting processes based on a non-normalised version of the empirical occupation measure~\cite{Cranston1995, Raimond1997} and discrete-time reinforced random walks; we refer the reader to~\cite{Pemantle2007} for a thorough historical account of self-interacting random processes. Asymptotic properties of SIDs have been studied in~\cite{Benaim2002,Benaim2003,Benaim2005,Benaim2011} restricting to compact spaces and symmetric interaction potentials and extended to $\mathbb{R}^d$ in~\cite{Kurtzmann2010,Chambeu2011,Kleptsyn2012} requiring confining potentials. More recently, in~\cite{Aleksian2022,Aleksian2024}, refinements of Kramer's law for SIDs have also appeared (as well as in~\cite{Barbier-Chebbah2024} for generalised overdamped Langevin systems). 

In parallel, various forms of SIDs have been employed in statistical physics to model autochemotaxis---the ability of organisms to communicate through local secretions that form chemical trails, which would enter $\rho_t$. Researchers have studied models where the signaling molecule remains in place to mark the trail, as seen in ants~\cite{Kranz2016,Kranz2019,DAlessandro2021,Barbier-Chebbah2022}, as well as models where the signal diffuses away, similar to bacteria~\cite{Tsori2004,Sengupta2009}. For the latter, stochastic modeling has provided insights into the behaviour of living microorganisms and self-propelled colloids, extending beyond traditional deterministic Keller--Segel equations~\cite{Brenner1998,Grima2005,Taktikos2012,Golestanian2012,Pohl2014,Gelimson2015,Grafke2017,Hokmabad2022}.

Most existing studies focus on the asymptotic properties of SIDs and autochemotactic systems, explaining typical behaviour. However, there remains a significant gap in our understanding of atypical and rare events in these systems, despite their crucial influence on future dynamics. To gain insights into the dynamics of SIDs, a theory of fluctuations and large deviations of time-averaged quantities, such as \eqref{eq:CurrObs}, is necessary. It is known, for instance, that long-range memory effects profoundly impact fluctuation behaviour, diverging from the established large deviation theory for Markovian cases~\cite{DenHollander2000,Touchette2009,Chetrite2015,Jack2020a,Carugno2022}. Non-Markovian random walks have revealed memory-driven effects~\cite{Schutz2004,Rebenshtok2007,Harris2009,Harris2015,Jack2020}, with studies showing a different \textit{speed} in the occurrence of large deviations. Recently, a long-time large deviation theory for a discrete-time version of an SID has been derived in~\cite{Budhiraja2022,Budhiraja2023}.

In this work, we provide an in-depth study of fluctuations of the observable \eqref{eq:CurrObs} using large deviation theory. In particular,
\begin{itemize}
\item In Sec.\ \ref{sec:Model} we detail the model and give an overview of its typical, or asymptotic, behaviour based on~\cite{Benaim2002}. We (re)derive the phase diagram for the SID using methods familiar to statistical phycisists, revealing a delocalisation-localisation phase transition from a regime where the SID is self-repelled to a regime where it is self-attracted by its own trajectory. \newtext{We have chosen to review these results to offer statistical physicists a single, concise paper where they can find the relevant information on SIDs in a language more familiar to them. All relevant literature is cited, allowing readers to access the original papers through the bibliography.}

\item In Sec.\ \ref{sec:methods}, we introduce two methodologies to analyse the fluctuating dynamics of \eqref{eq:CurrObs}, namely, (i) the \textit{adiabatic} approximation, where the system reaches its stationary distribution before current fluctuations develop, and (ii) an original extension of the so-called \textit{level 2.5 large deviations} for Markov processes. \newtext{The first method is based on a physical argument and is reminiscent of the adiabatic approximation that leads to the \textit{temporal additivity principle} discussed in~\cite{Harris2009}, although it differs in several aspects and in how it leads to a large deviation principle. We will provide more details on this in the relevant section below. For now, the essence of our adiabatic approximation lies in the well-behaved convergence of the empirical occupation measure to a fixed point, followed by the use of the homogeneous diffusion generator associated with this fixed point, as demonstrated in~\cite{Benaim2002}, to generate current fluctuations. While this first method is grounded in a physical argument, it is less `controllable' compared to the second method, which is based on formal calculation. The second method is original and not derived from previous publications, though similar equations have recently appeared in~\cite{Budhiraja2023} in the context of discrete-time self-interacting chains. In the following, we combine this method with perturbation theory to explicitly analyse current fluctuations.}

\item By comparison with Monte Carlo simulations, \newtext{we show in Sec.\ \ref{sec:results} that both methods provide  lower bounds on current fluctuations, which are tighter the less localised the SID is}. Additionally, we also demonstrate that both methods accurately estimate the asymptotic variance and suggest a phase transition at the onset of the localised regime.

\item Finally, in Sec.\ \ref{sec:conclusion}, we summarise our findings and outline open questions for future research.
\end{itemize}

\section{Model and typical behaviour}

\label{sec:Model}

We consider the general model for an SID introduced by B\"{e}naim, Ledoux and Raimond in~\cite{Benaim2002}, and then we restrict our focus to the specific SDE displayed in \eqref{eq:SIDEmp}. Subsequently, by using methods more familiar to the statistical physics community, we review the asymptotic properties of \eqref{eq:SIDEmp} and derive a phase diagram that characterises the long-term behaviour of the empirical occupation measure $\rho_t$.

\subsection{Model formulation}

We consider the general process $( X_t )_{t \geq 0}$, with $X_t \in \mathcal{M}$ a compact subset of $\mathbb{R}^d$, solution of the following SDE:
\begin{equation}
\label{eq:SIDGeneral}
d X_t = \sigma dW_t - \frac{1}{t} \left( \int_0^t \nabla V_{X_s}(X_t) ds \right) dt,
\end{equation}
where $W_t = \int_0^t dW_s$ is a Brownian motion on $\mathcal{M}$ characterised by independent and Gausssian increments, i.e., $\mathbb{E}[dW_t]=0$ and $\mathbb{E}[dW_t dW_{t'}] = \delta(t-t') dt$, $\sigma: \mathcal{M} \times \mathcal{M} \rightarrow \mathcal{M}$ a matrix that defines the coupling of the system to Gaussian white noise, and $V_x: \mathcal{M} \times \mathcal{M} \rightarrow \mathbb{R}$ a `potential' function responsible for the interaction between the current state of the system $X_t$ and all previous states $X_s$ for $0 < s < t$.

By introducing the empirical occupation measure
\begin{equation}
\label{eq:EmpOcc}
\rho_t(x) = \frac{1}{t} \int_0^t \delta(X_s - x) \, ds \, ,
\end{equation}
which is the fraction of time spent by the process on every position of the space $\mathcal{M}$, \eqref{eq:SIDGeneral} can be rewritten as 
\begin{equation}
\label{eq:SIDGenEmp}
d X_t = \sigma dW_t - \left( \int_{\mathcal{M}} \nabla V_{x}(X_t) \rho_t(x) \, dx \right) dt \, ,
\end{equation}
where the time integral has been replaced by a spatial integral over $\mathcal{M}$, clearly demonstrating the functional dependence of the process on its own trajectory via the empirical occupation measure $\rho_t$. For simplicity, we will refer to the entire process $( X_t )_{t \geq 0}$ as $X_t$, with context providing clarity. 

$X_t$ alone is not a Markov process, as long-range interactions in time directly enter the SDE and may drive the long-term behaviour of the system. We also introduce the notation
\begin{equation}
\label{eq:DriftRho}
F_{\rho_t}(X_t) = - \int_{\mathcal{M}} \nabla V_{x}(X_t) \rho_t(x) \, dx \, ,
\end{equation}
for the effective drift acting on the SID, which will be useful later on.

We will come back to the this general framework in Sec.\ \ref{sec:methods} when discussing methods to study fluctuations for SIDs, but for now we restrict ourselves to the case where $\mathcal{M} \equiv [0,2\pi)$, i.e., the unidimensional torus or ring with periodic boundary conditions, and consider the periodic potential
\begin{equation}
\label{eq:Potential}
V_{\theta}(\theta_t) = 2 c \cos(\theta - \theta_t + \phi) \, ,
\end{equation}
with $c \in \mathbb{R}^+$, $\phi \in [0,2\pi)$ and further restrict to $\sigma = \sqrt{D}$ with $D \in \mathbb{R}^+$ (also notice the change $X_t \rightarrow \theta_t$). We therefore rewrite \eqref{eq:SIDGenEmp} as
\begin{equation}
\label{eq:SIDEmpPot}
d \theta_t = \sigma dW_t - \left( \int_0^{2 \pi} \frac{d}{d\theta_t} 2 c \cos(\theta - \theta_t + \phi) \rho_t(\theta) \, d \theta \right) dt \, ,
\end{equation}
with the drift in \eqref{eq:DriftRho} given by the explicit form
\begin{equation}
\label{eq:DriftRhoRing}
F_{\rho_t}(\theta_t) = - \int_0^{2 \pi} \frac{d}{d\theta_t} 2 c \cos(\theta - \theta_t + \phi) \rho_t(\theta) \, d \theta  \, .
\end{equation}

The constant $c$ determines the strenght of the potential, the greater the $c$ the stronger the interaction between the current state $\theta_t$ and all previously visited states in $\rho_t$. The phase $\phi$, on the other hand, determines the type of potential. Consider the diffusion that originates in $\theta_0 = 0$. If $\phi = 0$, the potential is self-repelling because the particle starts from its maximum and is therefore attracted to the bottom of the potential, which lies at a distance $\pi$ from the origin. The opposite occurs when $\phi=\pi$; in this case, the particle originates already at the bottom of the potential and is most self-attracted. All intermediate cases can be explored by properly tuning $\phi$. It is also important to note that the interaction potential is integrated over time, and therefore, the effective drift experienced by the diffusing particle changes over time and is ultimately determined by the random realisation of the noise in \eqref{eq:SIDEmpPot}.

\subsection{Typical behaviour}

General results for the asymptotic behaviour of \eqref{eq:SIDGenEmp} for various classes of potentials have been investigated in~\cite{Benaim2002,Benaim2003,Benaim2005} with $\mathcal{M}$ considered as a general compact manifold using stochastic approximation methods~\cite{Borkar1998,Benaim1999}. These methods have been used to describe the asymptotic behaviour of the occupation measure $\rho_t$ by a limiting nonautonomous differential equation (see \eqref{eq:SIDNonautonomous} below). These results have later been extended to $\mathbb{R}^d$ in~\cite{Kurtzmann2010,Chambeu2011,Kleptsyn2012} for strictly convex interaction potentials. Remarkably, \cite{Benaim2005} showed that for symmetric potentials, $\rho_t$ almost surely converges to a local minimum of a nonlinear free-energy functional, which has many critical points, each chosen randomly with a certain probability. Specifically, when the potential is symmetric and self-repelling (known as Mercer potentials in the mathematics community), the free-energy functional is strictly convex, ensuring $\rho_t$ almost surely reaches a single global minimum in the long-time.

\subsubsection{Stationary Fokker--Planck equation}

Restricting to the case of \eqref{eq:SIDEmpPot} and assuming stationarity, the long-time behaviour $\rho_t \rightarrow \rho_{\text{inv}}$ can also be studied solving the following stationary non-linear Fokker--Planck equation:
\begin{equation}
\label{eq:SIDStatFokkerPlanck}
0 = \newtext{-} \frac{d}{d \theta} \left( F_{\rho_{\text{inv}}}(\theta) \rho_{\text{inv}}(\theta) \right) + \frac{D}{2} \frac{d^2 \rho_{\text{inv}}(\theta)}{d \theta^2} \, ,
\end{equation}
with the drift given by the spatial convolution
\begin{equation}
\label{eq:SIDStatFokkerPlanckAsymptDrift}
F_{\rho_{\text{inv}}}(\theta) = \newtext{-} \left( \int_0^{2 \pi} 2 c \sin(\theta' - \theta + \phi) \rho_{\text{inv}}(\theta') \, d \theta' \right) \, .
\end{equation}
We observe that a time-dependent Fokker–Planck equation for the stochastic process $(\theta_t)_{t \geq 0}$ alone is not defined, as the transition probability function necessarily depends on the occupation measure $\rho_t$. \newtext{Therefore, in Appendix \ref{appendix:FokkerPlanck}, we derive \eqref{eq:SIDStatFokkerPlanck} from a higher-dimensional, time-dependent Fokker–Planck equation associated with both $(\theta_t)_{t \geq 0}$ and $(\rho_t)_{t \geq 0}$. In fact, due to the explicit form of the drift \eqref{eq:DriftRhoRing}, the extended state space hosting the Markov process $(\theta_t, \rho_t)_{t \geq 0}$ is only three-dimensional. Starting from this time-dependent Fokker–Planck equation on the extended state space, \eqref{eq:SIDStatFokkerPlanck} is obtained as a long-time asymptotic result. This is, effectively, equivalent to assuming} that the effective drift experienced by the particle converges, \newtext{with} $\rho_t \rightarrow \rho_{\text{inv}}$. \newtext{Thus, we can consider} the stochastic process $(\theta_t)_{t \rightarrow \infty}$ as a simple Markov process whose drift is given by \eqref{eq:SIDStatFokkerPlanckAsymptDrift}. In this limit, \eqref{eq:SIDStatFokkerPlanck} is a well defined object to study, but it will not necessarily explain all asymptotic dynamics of \eqref{eq:SIDEmp}, such as the non-stationary asymptotic evolution of $\rho_t$.

A solution of \eqref{eq:SIDStatFokkerPlanck} is clearly given by the \textit{delocalised} function $\rho_{\text{inv}}(\theta) = (2 \pi)^{-1}$, representing the uniform distribution over the ring. However, this is not the only solution; another \textit{localised} solution also exists. By integrating \eqref{eq:SIDStatFokkerPlanck} once we obtain
\begin{equation}
\label{eq:SIDStatFokkerPlanck1}
J_{\rho_{\text{inv}}} = F_{\rho_{\text{inv}}}(\theta) \rho_{\text{inv}}(\theta) \newtext{-} \frac{D}{2} \frac{d \rho_{\text{inv}}(\theta)}{d \theta} \, ,
\end{equation}
where $J_{\rho_{\text{inv}}}$ is a constant expressing the stationary probability current of the process. A precise form of $J_{\rho_{\text{inv}}}$ can be obtained by imposing periodicity of the solution, viz.\ $\rho_{\text{inv}}(0)=\rho_{\text{inv}}(2\pi)$, and reads
\begin{equation}
\label{eq:SIDStatFokkerPlanckCurrent}
J_{\rho_{\text{inv}}} = \frac{c \left( \alpha_1^2 + \alpha_2^2 \right) \sin \phi}{\pi} \, ,
\end{equation}
where 
\begin{align}
\label{eq:SIDStatFokkerPlanckAlpha1}
\alpha_1 &= \int_0^{2 \pi} \rho_{\text{inv}}(\theta) \cos \theta \, d\theta \\
\label{eq:SIDStatFokkerPlanckAlpha2}
\alpha_2 &= \int_0^{2 \pi} \rho_{\text{inv}}(\theta) \sin \theta \, d\theta \, .
\end{align}
Noticeably, we expect that if a stationary solution $\rho_{\text{inv}}$ exists then $J_{\rho_{\text{inv}}} = 0$ because the long-time drift $F_{\rho_{\text{inv}}}$ in \eqref{eq:SIDStatFokkerPlanckAsymptDrift} is a conservative potential over $[0,2\pi)$. This condition over \eqref{eq:SIDStatFokkerPlanckCurrent} implies either $\phi = 0$ or $\phi = \pi$ for all $\alpha_1$ and $\alpha_2$ satisfying \eqref{eq:SIDStatFokkerPlanckAlpha1} and \eqref{eq:SIDStatFokkerPlanckAlpha2}, or $\alpha_1 = \alpha_2 = 0$ for all $\phi \in [0,2\pi)$, which results in the trivial uniform solution $\rho_{\text{inv}}(\theta) = (2 \pi)^{-1}$.

Equation \eqref{eq:SIDStatFokkerPlanck1} for $J_{\rho_{\text{inv}}} = 0$ can be rewritten as
\begin{equation}
\label{eq:SIDStatFokkerPlanck2}
\frac{D}{2} \frac{d \ln \rho_{\text{inv}}(\theta)}{d \theta} = - F_{\rho_{\text{inv}}}(\theta) \, ,
\end{equation}
and formally solved considering \eqref{eq:SIDStatFokkerPlanckAlpha1} and \eqref{eq:SIDStatFokkerPlanckAlpha2} via direct integration. The solution is
\begin{equation}
\label{eq:SIDStatFokkerPlanckSolution}
\begin{split}
\rho_{\text{inv}}(\theta) &= \frac{e^{-\frac{2}{D} \int_0^\theta F_{\rho_{\text{inv}}}(\theta') \, d\theta'}}{\mathcal{Z}} \\
&= \frac{e^{- \frac{4 c}{D} \left( \cos (\theta - \phi) \alpha_1 - \sin (\theta - \phi) \alpha_2 \right)}}{\mathcal{Z}} \, ,
\end{split}
\end{equation}
with $\mathcal{Z}$ the normalisation constant. Equation \eqref{eq:SIDStatFokkerPlanckSolution}, along with \eqref{eq:SIDStatFokkerPlanckAlpha1} and \eqref{eq:SIDStatFokkerPlanckAlpha2}, form a closed system of equations that can be solved as a function of $c$, $D$, and $\phi$. Summarising, we have found the following solutions 
\begin{equation}
\label{eq:SIDStatFokkerPlanckSolutionSummary}
\rho_{\text{inv}}(\theta) = 
\begin{cases}
(2 \pi)^{-1} &\hspace{1cm} \forall \; \phi \in [0,2\pi) \\
(\mathcal{Z})^{-1} e^{- \frac{4 c}{D} \left( \cos (\theta - \phi) \alpha_1 - \sin (\theta - \phi) \alpha_2 \right)} &\hspace{1cm} \text{for} \; \phi \in \left\lbrace 0, \pi \right\rbrace 
\end{cases} \, .
\end{equation}

These stationary solutions are not new. They already appeared in~\cite{Benaim2002} and were used to define a deterministic flow of a non-autonomous differential equation of the form
\begin{equation}
\label{eq:SIDNonautonomous}
\frac{d \rho_t}{d t} = - \rho_t + \frac{e^{-\frac{2}{D} \int_0^\theta F_{\rho_t}(\theta') \, d\theta'}}{\mathcal{Z}} \, ,
\end{equation}
to study the asymptotic behaviour of $\rho_t$. In the following, we discuss the stability of the solution $\rho_{\text{inv}}(\theta) = (2 \pi)^{-1}$, which will help draw a phase diagram for $\rho_{\text{inv}}$ and discuss an ergodicity-breaking phase transition controlled by the parameter $c/D$.

\subsubsection{Phase diagram for $\rho_{\text{inv}}$}
\label{subsubsec:PhaseDiagram}

\begin{figure}[ht]
    \centering
    \begin{subfigure}[b]{0.49\textwidth}
        \centering
        \includegraphics[width=\textwidth]{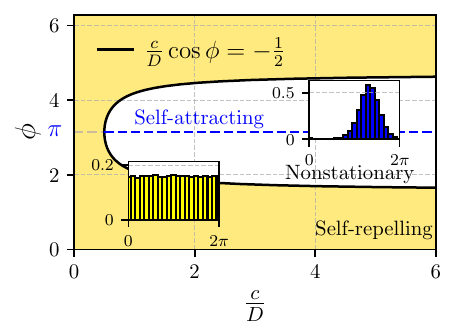}
    \end{subfigure}
    \hfill
    \begin{subfigure}[b]{0.49\textwidth}
        \centering
        \includegraphics[width=\textwidth]{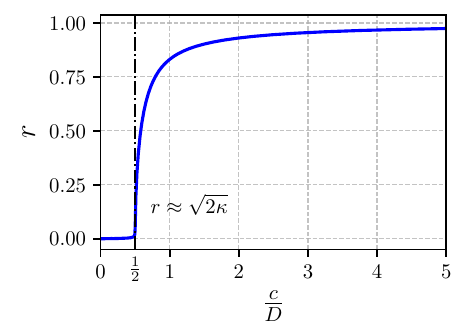}
    \end{subfigure}
    \caption{On the left, the phase diagram of $\rho_{\text{inv}}$ shows three distinguished phases: the self-repelling phase in yellow where the uniform stationary distribution is stable, the self-attracting phase as a blue dashed line where a new localised stationary distribution is stable and the uniform one is no longer and a white region where nonstationary behaviour is expected. \newtext{The histograms in the insets are typical long-time distributions of the SID obtained simulating \eqref{eq:SIDEmpPot} (see Sec.\ \ref{sec:results} and Appendix \ref{appendix:MonteCarlo} for details on simulations) for $c/D=0.01$ with $\phi=0$ (yellow) and $c/D=0.8$ with $\phi=\pi$ (blue)}.  On the right, the order parameter $r = \sqrt{\alpha_1^2 + \alpha_2^2}$ is plotted as a function of the control parameter $c/D$. As discussed in the text, $c/D=1/2$ marks a second-order phase transition characterised by a square-root singularity (see \eqref{eq:SquareRoot}), transitioning from the self-repelling phase to a self-attracting phase where the SID is typically localised.}
    \label{fig:typical}
\end{figure}

We show in the left panel of Fig.\ \ref{fig:typical} the phase diagram of $\rho_{\text{inv}}$. The region where the uniform solution \newtext{$\rho_{\text{inv}}(\theta) = (2\pi)^{-1}$} (first line of \eqref{eq:SIDStatFokkerPlanckSolutionSummary}) is stable is coloured in yellow and named the \textit{self-repelling} region as the SID is repelled by its own trajectory. 
\newtext{This region and its boundaries can be proved via a linear stability approach (see Appendix \ref{appendix:linear} and~\cite{Crawford1994,Strogatz1991,Strogatz2000}), which yields that the uniform solution is stable for
\begin{equation}
\label{eq:SIDStabilityErgodic}
\frac{c}{D} \cos \phi \geq - \frac{1}{2} \, .
\end{equation}}
The blue dashed line marks the region where the localised solution, second line of \eqref{eq:SIDStatFokkerPlanckSolutionSummary}, appears and is stable; this region is named \textit{self-attracting} because the SID is attracted by its own past and corresponds to a \textit{non-ergodic} phase for reasons that will become clear below. 

\newtext{The terms `self-attracting' and `self-repelling' were first introduced in~\cite{Benaim2002} in the context of SIDs. They indicate whether the \textit{attraction} is strong enough, or not, to overcome the Brownian motion. In other words, when the system is self-repelling, it does not necessarily mean it is repelled by itself, but rather that the attraction may not be sufficient to counteract the Brownian motion. This is particularly evident for $\phi = \pi$, where the process is always self-attractive, but for $c/D < 1/2$, the attraction is simply too weak. In the following, we will maintain this terminology, but readers should be aware of its intended meaning.}

The third white region cannot be understood by studying the stationary Fokker--Planck equation in \eqref{eq:SIDStatFokkerPlanck}, and therefore, it is where we expect non-stationary behaviour. We remark that for $\phi = 0$ the uniform distribution is always stable because of the positivity of both $c$ and $D$ whereas for $\phi = \pi$, at the onset $c/D = 1/2$, $\rho_{\text{inv}}(\theta) = (2\pi)^{-1}$ loses stability in favour of $\rho_{\text{inv}}(\theta) = (\mathcal{Z})^{-1} e^{- \frac{4 c}{D} \left( \cos (\theta - \phi) \alpha_1 - \sin (\theta - \phi) \alpha_2 \right)}$. In particular, in the following we show that the latter is not unique but come with an $O(2)$ symmetry, i.e., it is valid for all $\alpha_1$ and $\alpha_2$ such that $\alpha_1^2 + \alpha_2^2 = r^2$ where $r$ is a parameter that only depends on the ratio $c/D$ (see right panel of Fig.\ \ref{fig:typical}). Because of this property, the critical value $c/D = 1/2$ marks an ergodicity-breaking phase transition, viz.\ the invariant measure of the process is no longer unique for $c/D \geq 1/2$.

We consider $\rho_{\text{inv}}(\theta) = (\mathcal{Z})^{-1} e^{- \frac{4 c}{D} \left( \cos (\theta - \phi) \alpha_1 - \sin (\theta - \phi) \alpha_2 \right)}$. We remark that $\alpha_1$ and $\alpha_2$ fully determine where the typical value $\theta_{\text{max}}$ (the maximum) of the localised solution falls on the ring via 
\begin{equation}
\label{eq:Localisation}
\alpha_1 \tan (\theta_{\text{max}} - \phi) = - \alpha_2 \, .
\end{equation}
By moving to polar coordinates $\alpha_1 = r \cos \psi$ and $\alpha_2 = r \sin \psi$, \eqref{eq:SIDStatFokkerPlanckAlpha1} and \eqref{eq:SIDStatFokkerPlanckAlpha2} take the form 
\begin{align}
\label{eq:SIDAlpha1Polar}
r \cos \psi &= \int_0^{2 \pi} \cos \theta \frac{e^{-\frac{4 c r}{D} \left( \cos  (\psi - \phi + \theta) \right)}}{\mathcal{Z}} \, d \theta \\
\label{eq:SIDAlpha2Polar}
r \sin \psi &= \int_0^{2 \pi} \sin \theta \frac{e^{-\frac{4 c r}{D} \left( \cos  (\psi - \phi + \theta) \right)}}{\mathcal{Z}} \, d \theta \, .
\end{align}
\newtext{We now consider \eqref{eq:SIDAlpha1Polar} (the following steps can equivalently be performed using \eqref{eq:SIDAlpha2Polar}), introduce the change of variable $\psi - \phi + \theta = \theta'$ and apply the addition formula for the cosine to obtain
\begin{equation}
\label{eq:O2SymmetryCalc}
\begin{split}
r \cos \psi &= \cos \psi \int_0^{2 \pi} \cos (\phi + \theta') \frac{e^{-\frac{4 c r}{D}\cos \theta'}}{\mathcal{Z}} \, d\theta' + \sin \psi \int_0^{2 \pi} \sin (\phi + \theta') \frac{e^{-\frac{4 c r}{D}\cos \theta'}}{\mathcal{Z}} \, d\theta'  \\
&= \cos \psi \int_0^{2 \pi} \cos (\phi + \theta') \frac{e^{-\frac{4 c r}{D}\cos \theta'}}{\mathcal{Z}} \, d\theta' \, ,
\end{split}
\end{equation}
where the integral over the sine function disapper for symmetry reasons as $\phi \in \left\lbrace 0, \pi \right\rbrace$. Also notice that the second line in \eqref{eq:O2SymmetryCalc} is valid for all $\psi \in [0,2\pi)$ or, in the equivalent cartesian formulation, $\alpha_1^2 + \alpha_2^2 = r^2$, and therefore
\begin{equation}
\label{eq:O2Symmetry}
r = \int_0^{2 \pi} \cos (\phi + \theta') \frac{e^{-\frac{4 c r}{D}\cos \theta'}}{\mathcal{Z}} \, d\theta' \, .
\end{equation}}

In the following, we calculate the value of the parameter $r$. For $\phi=0$, \eqref{eq:O2Symmetry} can be written in terms of $I_0$ and $I_1$, i.e., the modified Bessel functions of the first kind~\cite{Boisvert2011}, as
\begin{equation}
\label{eq:O2SymmetryPhi0}
r = \int_0^{2 \pi} \cos \theta' \frac{e^{-\frac{4 c r}{D}\cos \theta'}}{\mathcal{Z}} \, d\theta' = - \frac{I_1 \left( \frac{4 c r}{D} \right)}{I_0 \left( \frac{4 c r}{D} \right)} \, .
\end{equation}
In particular, as $I_0 \geq 0$ and $I_1 \geq 0$ for $r \geq 0$, the only solution to \eqref{eq:O2SymmetryPhi0} is $r=0$ and therefore $\alpha_1=\alpha_2=0$: we re-obtain the stable uniform distribution. On the other hand, for $\phi=\pi$, \eqref{eq:O2Symmetry} takes the form
\begin{equation}
\label{eq:O2SymmetryPhiPi}
r = - \int_0^{2 \pi} \cos \theta' \frac{e^{-\frac{4 c r}{D}\cos \theta'}}{\mathcal{Z}} \, d\theta' = \frac{I_1 \left( \frac{4 c r}{D} \right)}{I_0 \left( \frac{4 c r}{D} \right)} \, ,
\end{equation}
and in this case it is evident that a solution $r > 0$ is admitted. The value of $r$ can be calculated numerically by solving \eqref{eq:O2SymmetryPhiPi} and depends solely of the ratio $c/D$. We plot the behaviour of $r$ as a function of $c/D$ in the right panel of Fig.\ \ref{fig:typical}. Evidently, the greater the ratio (e.g., the smaller the noise), the higher the value of $r$ which asymptotically tends to $1$. Conversely, for very small ratios (e.g., high noise), $r$ approaches $0$.

The critical value $c/D = 1/2$ that is marked in Fig.\ \ref{fig:typical} can be derived from \eqref{eq:O2SymmetryPhiPi} by taking a derivative with respect to $r$ at $r=0$ on both sides of the equation and imposing that left and right hand side of the equation must be the same. If we do so, we get
\begin{equation}
1 = \frac{d}{d r}\at[\Big]{r=0} \left( \frac{I_1(k r)}{I_0(k r)} \right) = \frac{k}{2} + \frac{k}{2} \frac{I_2(k r)}{ I_0(k r)}\at[\Big]{r=0} - k \frac{I_1^2(k r)}{I_0^2(k r)}\at[\Big]{r=0} = \frac{k}{2} \, ,
\end{equation}
where we use the shorthand notation $k = 4c/D$ and the last equality follows from properties of the modified Bessel functions of the first kind~\cite{Boisvert2011}. Evidently, we must have $k = k_c \coloneqq 2$ and therefore $c/D = 1/2$ at criticality. By expanding \eqref{eq:O2SymmetryPhiPi} for $r \rightarrow 0$ and keeping the lowest meaningful order in $r$ we also get
\begin{equation}
\label{eq:SquareRoot}
r = \sqrt{\frac{8 (k - 2)}{k^3}} \stackrel{k \approx k_c}{\approx} \left( 2 \kappa \right)^\beta \, ,
\end{equation}
with $\kappa = (k-k_c)/k_c$ and $\beta = 1/2$. 

In conclusion, $k_c = 2$ or, equivalently, $c/D = 1/2$ marks a second-order phase transition characterised by a square-root singularity from a self-repelling phase for $c/D < 1/2$ where the only stationary solution of the SID is uniform over the ring to a self-attracting phase where the SID typically localises at $\theta_{\text{max}}$ solution of \eqref{eq:Localisation} with an $O(2)$ symmetry. Given the fact that where the process localises depends on the realisation of the noise and therefore the SID admits more than one stationary distribution, the self-attracing phase is non-ergodic.

This phase transition is reminiscent of a disorder-order phase transition of a mean-field XY model (also known as mean-field Kuramoto or Sakaguchi model). In fact, the asymptotic behaviour is the same and, in the Appendix \ref{appendix:XY}, we show that the stationary Dean's equation of a mean-field XY model, in the limit of infinitely many interacting spins, reduces to the stationary non-linear Fokker-Planck equation in \eqref{eq:SIDStatFokkerPlanck}. Notably, works in this field, such as~\cite{Sakaguchi1988,Strogatz1991,Strogatz2000,Crawford1994} use a parameter very similar to $r$ in \eqref{eq:O2Symmetry}, representing the phase-overlap responsible for the syncronisation among all spins in the lattice. Similarly, here $r$ can be interpreted to represent the `synchronisation' of the SID with its own past trajectory.

The parallel between the SID of \eqref{eq:SIDGenEmp} and interacting many-particle systems \newtext{seems to be} more profound, extending as well to the much-studied McKean-Vlasov process in probability theory. The McKean-Vlasov process is a stochastic process described by an SDE whose drift depends on the law of the process itself, i.e., the probability distribution of the process in the configuration space. Under certain assumptions, such as a self-interacting potential, the law of a McKean--Vlasov process converges to the invariant measure $\rho_\text{inv}$, the solution of \eqref{eq:SIDStatFokkerPlanck} (see~\cite{Cattiaux2008,Kleptsyn2012,AleksianThesis2023}). \newtext{This suggests that} the asymptotic properties of these two processes \newtext{could} be studied similarly. For instance, it is customary in probability to study the typical behaviour of a McKean--Vlasov process by examining the behaviour of a many-particle system with mean-field interactions~\cite{MalrieuThesis2001}. \newtext{It would be interesting to explore whether a similar approach could be used to study the typical behaviour and, potentially, the typical fluctuations of an SID. Under certain assumptions, it has been shown that its empirical occupation measure converges to a critical point of a free energy functional~\cite{Benaim2005}, which could provide a basis for mean-field-like techniques.}

\subsection{Current observable}

In the following, we focus on the fluctuations of the current observable in \eqref{eq:CurrObs} for $\phi=\pi$. We will move along the blue dashed line in Fig.\ \ref{fig:typical} by varying the ratio $c/D$ and study how the fluctuation behaviour of $\Omega_t$ changes as we transition from the self-repelling region, characterised by a delocalised solution for $\rho_{\text{inv}}$, to the self-attracting region, where the solution $\rho_{\text{inv}}$ is localised. Although we only study the behaviour for $\phi=\pi$, the results found to hold in the self-repelling case extend to the whole self-repelling region of Fig.\ \ref{fig:typical}. This work does not focus on the case where $\rho_t$ does not admit a stationary solution (white region in the left panel of Fig.\ \ref{fig:typical}).

In the infinite-time limit we expect $\Omega_t$ to converge to the stationary probability current, or average speed, $J_{\rho_{\text{inv}}}$. This object, in the relevant case of having a stationary behaviour $\rho_t \rightarrow \rho_{\text{inv}}$, is shown above to be $J_{\rho_{\text{inv}}} = 0$ (see~\cite{Reimann2001,Pavlov2012} for general derivations in the case of a drifted particle in a periodic potential). Hence, we have
\begin{equation}
\label{eq:CurrSIDConcentration}
\Omega_t \rightarrow 0 \, ,
\end{equation}
for almost all (in the probability sense) paths of the SID in \eqref{eq:SIDEmp}. Although the current \textit{concentrates} at $0$ in the long time limit, $\Omega_t$ fluctuates around this typical value. It is this fluctuation behaviour that is of interest here.

According to large deviation theory, for a (possibly driven) diffusing particle on the ring subject to a periodic potential (not the SID in \eqref{eq:SIDEmp}), the distribution of $\Omega_t$ is shown in~\cite{TsobgniNyawo2016} to take the general form
\begin{equation}
\label{eq:LDDrivenParticle}
P(\Omega_t = \omega) = e^{- t I(\omega) + o(t)}
\end{equation}
in the limit $t \rightarrow \infty$. The dominant scaling term in \eqref{eq:LDDrivenParticle} is exponential and its \textit{rate function} $I$ is given by the limit 
\begin{equation}
\label{eq:RateFunction}
I(\omega) = - \lim_{t \rightarrow \infty} \frac{1}{t} \ln P(\Omega_t = \omega) \, .
\end{equation}
This object controls the rate at which the probability $P(\Omega_t = \omega)$ decays to $0$ with time for any $\omega \neq J_{\text{typ}}$. As a consequence, $I \geq 0$ with equality reached only at $\omega = J_{\text{typ}}$. Evidently, the fact that $I > 0$ for any other values of $\omega$ shows that fluctuations away from the typical value are exponentially unlikely at large times.

Although the long-time limit of the current $\Omega_t$ is understood for the SID in \eqref{eq:SIDEmp} as given in \eqref{eq:CurrSIDConcentration}, its fluctuation behaviour has not yet been proved to satisfy the large deviation relation in \eqref{eq:LDDrivenParticle}. In the following, we provide evidence that such a large deviation behaviour may hold for the SID, at least under certain assumptions.


\section{Methods}
\label{sec:methods}

In this Section, we discuss two methods used in Sec.\ \ref{sec:results} to study the fluctuations of the current observable \eqref{eq:CurrObs}. The first method, adiabatic approximation, is based on a separation of time scales, an approach commonly used both in probability theory and statistical physics. The second method is new and is based on an extension of level 2.5 of large deviation theory to non-Markov processes of the general form \eqref{eq:SIDGeneral}.

\newtext{Other approaches to study fluctuations of \eqref{eq:CurrObs} could be considered, such as extending the state space---similar to Appendix \ref{appendix:FokkerPlanck}---by including the dynamics of $\Omega_t$ as well to make the system Markovian and analysing the time evolution of the associated Fokker--Planck equation. While this is a more rigorous and potentially interesting avenue, we have chosen not to pursue it due to the non-autonomous drift and the time-dependent, correlated noise matrix that would appear in the Fokker--Planck equation. These features are undoubtedly interesting and worth exploring but would require significant time to properly develop suitable methods.}

\subsection{Adiabatic approximation}
\label{subsec:adiabatic}

By simulating the SID process in \eqref{eq:SIDEmp} in the relevant region of the phase diagram in Fig.\ \ref{fig:typical}, i.e., $\phi = \pi$, we observe that the effective potential (and consequently the drift) converges rapidly in both the self-repelling and self-attracting phases. As we are interested in the long time fluctuations of the current, it seems reasonable to assume a separation of time scales whereby $\rho_t \rightarrow \rho_{\text{inv}}$, solution of \eqref{eq:SIDStatFokkerPlanck}, more rapidly than $\Omega_t \rightarrow J_{\rho_{\text{inv}}} (\equiv 0)$. 

\newtext{In~\cite{Harris2009}, a similar approach is used, where an adiabatic approximation assumes that current values do not change significantly over long time slices. In the long-time limit, this effectively disentangles currents from states, allowing currents to adapt to new state configurations in subsequent time slices. In our context, however, the disentangling occurs once and for all in the long-time limit, as we assume the empirical measure converges to a stationary distribution. As a result, we do not account for the initial transient regime, and the current fluctuations are not allowed to adapt to different distributions over long time slices.}

\newtext{We also note that using such an approximation does not necessarily require knowledge of all the time scales involved, which remain an interesting open question. This is because the adiabatic approximation assumes that the system has already relaxed to its stationary distribution, which must occur after all transient regimes, potentially characterised by strong non-Markovian behaviour. Additionally, it is relatively straightforward to compare this approximation with simulations, as these can simply be run for a sufficiently long time.}

The current in \eqref{eq:CurrObs} is now defined on the Markov process
\begin{equation}
\label{eq:SIDAdiabatic}
d \theta_t = \sqrt{D} dW_t + 2 c \left( \cos(\theta_t) \alpha_2 - \sin(\theta_t) \alpha_1  \right) dt \, ,
\end{equation}
which has been obtained from \eqref{eq:SIDEmp} replacing $\rho_t$ with $\rho_{\text{inv}}$ and where $\alpha_1$ and $\alpha_2$ are given by \eqref{eq:SIDStatFokkerPlanckAlpha1} and \eqref{eq:SIDStatFokkerPlanckAlpha2} respectively.

In the self-repelling case we have $\alpha_1 = \alpha_2 = 0$ and the process in \eqref{eq:SIDAdiabatic} reduces to a trivial Brownian motion. In the self-attracting case instead, $\alpha_1$ and $\alpha_2$ are not known a priori and their value depend on the realisation of the noise driving the SID. However, they simply are constant values and therefore we can study the fluctuation behaviour of $\Omega_t$ considering them as parameters.

As mentioned earlier, due to the Markovian character of \eqref{eq:SIDAdiabatic} it is already known that the fluctuations of $\Omega_t$ take the general large deviation form in \eqref{eq:LDDrivenParticle}. Therefore, we only need to calculate the rate function, namely $I_a$ where the subscript $a$ stands for adiabatic. 

To do so we use the G\"{a}rtner--Ellis theorem\newtext{~\cite{Touchette2009,Dembo2010,Touchette2018}}, which states that the rate function is given by the Legendre--Fenchel transform
\begin{equation}
\label{eq:LegendreFenchel}
I_a(\omega) = \sup_{k \in \mathbb{R}} \left( k \omega - \lambda_a(k) \right) \, ,
\end{equation}
of the scaled cumulant generating function (SCGF) of $\Omega_t$ given by 
\begin{equation}
\label{eq:SCGF}
\lambda_a(k) = \lim_{t \rightarrow \infty} \frac{1}{t} \ln \mathbb{E} \left[ e^{t k \Omega_t} \right] \, ,
\end{equation}
when the latter exists and is differentiable with respect to $k$. For time-additive observables of Markov processes, such as $\Omega_t$, $\lambda_a$ is known to be given by the dominant eigenvalue of the so-called tilted generator\newtext{~\cite{Chetrite2015,TsobgniNyawo2016,Touchette2018}}
\begin{equation}
\label{eq:TiltedGenerator}
\mathcal{L}_k = 2c \left( \cos(\theta_t) \alpha_2 - \sin(\theta_t) \alpha_1  \right) \left( \frac{d}{d \theta} + k \right) + \frac{D}{2} \left( \frac{d}{d \theta} + k \right)^2 \, ,
\end{equation}
which acts on bounded periodic functions of $[0,2\pi)$. $\mathcal{L}_k$ is a linear operator obtained by modifying/tilting the infinitesimal generator of \eqref{eq:SIDAdiabatic}. For further details on the theory we refer the reader to~\cite{Chetrite2015,Touchette2018} and for details related to a driven diffusive particle in a periodic potential to~\cite{TsobgniNyawo2016}.

The eigenvalue problem that we need to solve reads
\begin{equation}
\label{eq:EigProblem}
\mathcal{L}_k r_k = \lambda_a(k) r_k \, ,
\end{equation}
where $r_k$ is the dominant right eigenfunction of the tilted operator in \eqref{eq:TiltedGenerator}. We follow~\cite{TsobgniNyawo2016} and expand $r_k$ in a Bloch--Fourier series as
\begin{equation}
\label{eq:RightFourier}
r_k = \sum_{n=-\infty}^{\infty} c_n e^{i n \theta} .
\end{equation}
Replacing \eqref{eq:RightFourier} in \eqref{eq:EigProblem} along with \eqref{eq:TiltedGenerator}, using Euler formulas for cosine and sine, and performing some algebraic manipulations and sum shifts, we obtain the following tridiagonal system of recurrence relations:
\begin{equation}
\label{eq:TridCurr}
\begin{split}
&c_n \left( - \frac{D}{2}n^2 + k D i n + \frac{D}{2} k^2 - \lambda_a(k) \right) + c_{n-1} \left[ c \alpha_2 \left( i (n-1) + k \right) + c \alpha_1 \left(1-n + i k \right) \right] + \\
&\hspace{4cm} + c_{n+1} \left[ c \alpha_2 \left( i (n+1) + k \right) + c \alpha_1 \left(n+1 - i k \right) \right] = 0 \, .
\end{split}
\end{equation}
This can be solved by truncating to some finite value of $n$ and studying the kernel of the truncated tridiagonal matrix. The SCGF $\lambda_a$ is left as a parameter in the system and is determined as the real value for which the determinant of the matrix becomes zero. This condition is necessary and sufficient for having a non-trivial kernel. \newtext{We also note that truncating tridiagonal systems of the form \eqref{eq:TridCurr} is a standard approach for finding approximate solutions, and it is not an issue as long as the truncation value $n$ is chosen carefully. In general, truncating an infinite tridiagonal matrix can introduce boundary effects and alter the spectral properties, potentially distorting the kernel or misrepresenting null space vectors. To avoid such issues, we ensure convergence by gradually increasing the truncation size and have found that $n = 30$ offers a good balance between computational feasibility and accuracy (as shown in Figs.\ \ref{fig:asymptvar} and \ref{fig:monte} in Sec.\ IV).} 

We notice that in the simple self-repelling case, since \eqref{eq:SIDAdiabatic} reduces to a Brownian motion, we already know that fluctuations have a Gaussian shape and therefore
\begin{equation}
I_a(\omega) = \frac{\omega^2}{2 D} \, .
\end{equation}
In the self-attracting case instead, $I_a$ is a complicated function and it will be calculated with the method explained above. In Sec.\ \ref{sec:results} we will plot the rate functions $I_a$ obtained in the self-repelling and self-attracting case and compare them with simulations of the SID.

\subsection{Extended level 2.5 large deviations}
\label{subsec:extended}

\subsubsection{Markov processes}

Given a Markov diffusion process $( Y_t )_{t \geq 0}$, with $Y_t \in \mathbb{R}^d$, the level 2.5 of large deviations describes the likelihood of joint fluctuations of the empirical occupation measure $\rho_t$ in \eqref{eq:EmpOcc} and empirical current given by the formal relation
\begin{equation}
\label{eq:EmpCurr}
j_t(y) = \frac{1}{t} \int_0^t \delta(Y_t - y) \circ dY_t \, .
\end{equation}
where $\circ$ symbolises the use of Stratonovich integration. The joint distribution of empirical occupation and current $P(\rho_t = \rho, j_t = j)$ is known to take a large deviation form with an explicit rate function, namely $I_{\text{2.5}}$. \footnote{Intuitively, the form is explicit because for Markov processes drift and diffusion are uniquely identified by the stationary distribution and the stationary current of the process.} This function can be calculated in different ways, either by combining \textit{tilting} with the Girsanov relation\newtext{~\cite{Maes2008,Barato2015}} or using spectral methods combined with the G\"{a}rtner--Ellis theorem\newtext{~\cite{Barato2015}}. \newtext{See also~\cite{Chernyak2009}, for a field-theoretic approach.} Regardless of the method applied, the joint rate function is
\begin{equation}
\label{eq:Level2.5Markov}
I_{\text{2.5}}[\rho,j] = \begin{cases}
\frac{1}{2} \int_{\mathbb{R}^d} (j - J_{F,\rho})(\rho D)^{-1} (j - J_{F,\rho}) \, dy &\hspace{1cm} \text{if } \nabla \cdot j = 0 \\
\infty &\hspace{1cm} \text{otherwise} \, ,
\end{cases}
\end{equation}
where
\begin{equation}
\label{eq:ProbaCurrent2.5}
J_{F,\rho} = F \rho - \frac{D}{2} \nabla \rho \, ,
\end{equation}
is the probability current associated with the drift $F$ of the Markov process $Y_t$ and density $\rho$ and $D$ is, in general, any invertible noise matrix (not necessarily a scalar) given by $D = \sigma \sigma^\top$ entering \eqref{eq:SIDGeneral}.

\subsubsection{Self-interacting diffusions}
\label{subsubsec:SIDs}

In the following, we show a heuristic derivation of an extension of level 2.5 large deviations for SIDs with the general form \eqref{eq:SIDGeneral}. The validity of the formula obtained below in \eqref{eq:Level2.5SID} will be tested in Sec.\ \ref{sec:results}. In the calculation we apply a tilting approach. However, given that the noise appearing in \eqref{eq:SIDGeneral} is additive, we will not need to resort to the Girsanov relation. Furthermore, we work on the validity of two assumptions: 
\begin{itemize}

\item There exists an ergodic process $( Y_t )_{t \geq 0}$ with probability path measure $d \mathbb{P}_{Y_t, \left[ 0,t \right]}$, such that its empirical occupation measure $\rho_t$ converges to $\rho$ and its empirical current $j_t$ in \eqref{eq:EmpCurr} converges to $j \coloneqq J_{\tilde{F}_{\rho},\rho}$ where
\begin{equation}
\label{eq:DriftAuxiliary2.5}
\tilde{F}_{\rho}(x) = - \int_{\mathbb{R}^d} \nabla \tilde{V}_y(x) \rho(x) \, dx \, , 
\end{equation}
is the spatial convolution of $\tilde{V}_y$, an auxiliary self-interacting potential, and $\rho$ the invariant measure of $Y_t$. This auxiliary process is a (Markov) diffusion with noise matrix $D$, but with a modified drift given by \eqref{eq:DriftAuxiliary2.5} which can be cast as a function of $j$ as 
\begin{equation}
\label{eq:DriftAuxiliary2.5Current}
\tilde{F}_{\rho}(x) = \frac{j(x)}{\rho(x)} + \frac{D}{2} \nabla \ln \rho(x) \, .
\end{equation}
As a consequence, $\rho$ is unique and $\nabla \cdot j = 0$.
 
\item The original path measure of $( X_t )_{t \geq 0}$ and the auxiliary one of $( Y_t )_{t \geq 0}$ are absolutely continuous with respect to each other. This implies that the Radon--Nikodym derivative is well defined and its asymptotic behaviour is identified by a function $I^{\text{(SID)}}_{\text{2.5}}$ such that the relation 
\begin{equation}
\label{eq:RadonNikodim}
\frac{d \mathbb{P}_{X_t, \left[ 0,t \right]}}{d \mathbb{P}_{Y_t, \left[ 0,t \right]}}[x] = e^{- t I^{\text{(SID)}}_{\text{2.5}}[x] + o(t)} \, ,
\end{equation}
with $x$ a general path, is well defined.
\end{itemize}

These two assumptions already seem to impose strict limitations on the types of fluctuations we can analyse for SIDs. Notably, an SID is not a Markov process, yet we are assuming that a fluctuation in a non-Markov process can be represented by a Markov process, where the drift is uniquely generated by the spatial convolution of an auxiliary interaction potential and the stationary distribution $\rho$. At best, any information that the function $I^{\text{(SID)}}_{\text{2.5}}$ provides regarding the likelihood of fluctuations will only serve as an upper bound in terms of rate functions or a lower bound in terms of actual probabilities. We will investigate in Sec.\ \ref{sec:results} how effective this approach is in providing a bound.

We start off by formally re-expressing the joint probability of density and current in terms of the path measure of $X_t$, i.e.,
\begin{equation}
\label{eq:1DerivationLEvel2.5}
\begin{split}
P(\rho_t = \rho, j_t = j) &= \mathbb{E}_{X_t} \left[ \delta_{\rho_t[X_t],\rho} \delta_{j_t[X_t],j} \right] \\
&= \int d \mathbb{P}_{X_t, \left[ 0,t \right]}[x] \delta_{\rho_t[x],\rho} \delta_{j_t[x],j} \, ,
\end{split}
\end{equation}
and select paths $( X_t )_{t \geq 0} = x$ using Kronecker-$\delta$ (or indicator) functions. We now introduce the new tilted path measure $d \mathbb{P}_{Y_t, \left[ 0,t \right]}$ and express the newly-appearing Radon--Nikodim derivative as in \eqref{eq:RadonNikodim} to get
\begin{equation}
\label{eq:2DerivationLEvel2.5}
P(\rho_t = \rho, j_t = j) \sim \mathbb{E}_{Y_t} \left[ e^{- t I^{\text{(SID)}}_{\text{2.5}}[Y_t]} \delta_{\rho_t[Y_t],\rho} \delta_{j_t[Y_t],j} \right] \, .
\end{equation}
Eventually, noticing that $\mathbb{E}_{Y_t} \left[ \delta_{\rho_t[Y_t],\rho} \delta_{j_t[Y_t],j} \right] = 1$ because of the properties of the auxiliary process $Y_t$, we get
\begin{equation}
\label{eq:3DerivationLEvel2.5}
P(\rho_t = \rho, j_t = j) \sim e^{- t I^{\text{(SID)}}_{\text{2.5}}[\rho,j]} \, ,
\end{equation}
and we are only left with calculating $I^{\text{(SID)}}_{\text{2.5}}$. 

In the simple case of additive noise Onsager--Machlup fluctuation theory applies and we can therefore write
\begin{equation}
\label{eq:RadonNikodimOnsagerMachlup}
\ln \left( \frac{d \mathbb{P}_{X_t, \left[ 0,t \right]}}{d \mathbb{P}_{Y_t, \left[ 0,t \right]}}[X_t] \right) = \ln \left( \frac{\exp^{- \int_0^t ds \frac{1}{2} \left[ \left( \dot{X}_s - F_{\rho_s} \right)^\top \cdot D^{-1} \left( \dot{X}_s - F_{\rho_s} \right) + \frac{1}{2} \nabla \cdot F_{\rho_s} \right]}}{\exp^{- \int_0^t ds \frac{1}{2} \left[ \left( \dot{X}_s - \tilde{F}_{\rho} \right)^T \cdot D^{-1} \left( \dot{X}_s - \tilde{F}_{\rho} \right) + \frac{1}{2} \nabla \cdot \tilde{F}_{\rho} \right]}} \right) \, ,
\end{equation}
where we have used the It\^{o} convention and therefore divergence terms appear in the integrals and we mark with $\dot{}$ the temporal derivative. Furthermore, we notice that at the numerator the drift $F_{\rho_t}$ is defined as in \eqref{eq:DriftRho} and is non-Markovian in nature, whereas at the denominator $\tilde{F}_{\rho}$ is given in \eqref{eq:DriftAuxiliary2.5Current} and is related to the auxiliary Markov process introduced above. Expanding the products on the right hand side of \eqref{eq:RadonNikodimOnsagerMachlup}, merging numerator and denominator in a single expression and introducing the definitions of $\rho_t$ in \eqref{eq:EmpOcc} and $j_t$ in \eqref{eq:EmpCurr} we get
\begin{equation}
\label{eq:RadonNikodimOnsagerMachlup1}
\ln \left( \frac{d \mathbb{P}_{X_t, \left[ 0,t \right]}}{d \mathbb{P}_{Y_t, \left[ 0,t \right]}}[X_t] \right) = - t \int_{\mathbb{R}^d} j_t \cdot D^{-1} \left( \tilde{F}_{\rho} - F_{\rho_t} \right) \, dx - \frac{t}{2} \int_{\mathbb{R}^d} \rho_t \left[ \nabla \cdot  \left( F_{\rho_t} - \tilde{F}_{\rho} \right) - ( F_{\rho_t} - \tilde{F}_{\rho}) \cdot D^{-1} (F_{\rho_t} + \tilde{F}_{\rho}) \right] \, dx \, .
\end{equation}

By replacing $\tilde{F}_{\rho}$ from \eqref{eq:DriftAuxiliary2.5Current} and $J_{F_{\rho_t},\rho}$ from \eqref{eq:ProbaCurrent2.5}, with the change $F \ra F_{\rho_t}$, we get
\begin{equation}
\label{eq:RadonNikodimOnsagerMachlup2}
\begin{split}
&\ln \left( \frac{d \mathbb{P}_{X_t, \left[ 0,t \right]}}{d \mathbb{P}_{Y_t, \left[ 0,t \right]}}[X_t] \right) = \\
&\hspace{0.5cm} - t \int_{\mathbb{R}^d} \left\lbrace j_t \cdot D^{-1} \left( \frac{j - J_{F_{\rho_t},\rho}}{\rho}\right) + \frac{\rho_t}{2} \left[ \nabla \cdot  \left( \frac{-j + J_{F_{\rho_t},\rho}}{\rho} \right) - \left( \frac{j - J_{F_{\rho_t},\rho}}{\rho} \right) \cdot D^{-1} \left( \frac{j - J_{F_{\rho_t},\rho}}{\rho} - 2 \tilde{F}_{\rho} \right) \right] \right\rbrace \, dx \, .
\end{split}
\end{equation}

Finally, integrating by parts the second term in the integral, collecting the common factor and using again \eqref{eq:DriftAuxiliary2.5Current} to isolate a current $j$, we get
\begin{equation}
\label{eq:Level2.5SID}
I^{\text{(SID)}}_{\text{2.5}}[\rho,j] = \begin{cases}
\frac{1}{2} \int_{\mathbb{R}^d} (j - J_{F_{\rho},\rho}) \cdot(\rho D)^{-1} (j - J_{F_{\rho},\rho}) \, dx &\hspace{1cm} \text{if } \nabla \cdot j = 0 \\
\infty &\hspace{1cm} \text{otherwise} \, .
\end{cases}
\end{equation}
This rate function is similar to the one known for diffusion processes in \eqref{eq:Level2.5Markov}, with the only difference that the instantaneous current $J_{F_{\rho},\rho}$ now depends on a drift that is represented by a spatial convolution with the density $\rho$. \newtext{A similar rate function was recently considered in~\cite{Budhiraja2023} as an upper bound for discrete-time self-interacting Markov chains.}

For the case of interest in this paper, that is the SID in \eqref{eq:SIDEmp}, \eqref{eq:Level2.5SID} reduces to
\begin{equation}
\label{eq:Level2.5SIDRing}
I^{\text{(SID)}}_{\text{2.5}}[\rho,j] =
\begin{cases}
\frac{1}{2 D} \int_0^{2 \pi} \rho^{-1} J^2_{F_{\rho},\rho} \, d\theta + \frac{j^2}{2 D} \int_0^{2 \pi} \rho^{-1} \, d\theta - \frac{j}{D} \int_0^{2 \pi} \rho^{-1} J_{F_{\rho},\rho} \, d\theta &\hspace{1cm} \text{if } \frac{d j}{d \theta} = 0 \\
\infty &\hspace{1cm} \text{otherwise}
\end{cases} \, ,
\end{equation}
where it is evident that $j$ must be a constant over the ring. We can further simplify the expression in \eqref{eq:Level2.5SIDRing} expanding $J_{F_{\rho},\rho}$ and manipulating the expression appropriately to find
\begin{equation}
\label{eq:Level2.5SIDRingFinal}
I^{\text{(SID)}}_{\text{2.5}}[\rho,\omega] =
\begin{cases}
\frac{D}{2} \int_0^{2 \pi} \left[ \rho \left( \frac{\rho'}{2 \rho} - \frac{F_{\rho}}{D} \right)^2 + \frac{\omega^2}{4 \pi^2 D^2}  \rho^{-1} \right] \, d\theta &\hspace{1cm} \text{if } \frac{d \omega}{d \theta} = 0 \\
\infty &\hspace{1cm} \text{otherwise}
\end{cases} \, ,
\end{equation}
where $'$ denotes a spatial derivative, and we have used the fact that integration by parts gives $\int_0^{2 \pi} \rho^{-1} \rho' \, d\theta = 0$. We also consider that $\Omega_t$ in \eqref{eq:CurrObs} is related to $j_t$ via $\Omega_t = 2\pi j_t$, and therefore, $\omega = 2\pi j$.

\subsubsection{Current contraction and the Euler--Lagrange equation}

We are interested in the current large deviations. Given the formula for the extended level 2.5 rate function in \eqref{eq:Level2.5SIDRingFinal}, we can obtain the current rate function via contraction as follows:
\begin{equation}
\label{eq:InfimumRhoLevel2.5}
I(\omega) = \inf_{\substack{\rho \\ 1 = \int_0^{2 \pi} \rho}} I^{\text{(SID)}}_{\text{2.5}} \left[ \rho,\omega \right] \, ,
\end{equation}
where we notice that no minimisation has to be carried out over the current $\omega$ because the latter is a constant over the circle as remarked just below \eqref{eq:Level2.5SIDRing}.

The minimisation in \eqref{eq:InfimumRhoLevel2.5} can be tackled by introducing the Lagrangian functional
\begin{equation}
\label{eq:LagrLDCurr}
\mathcal{L}[\rho,\rho'] = \frac{D}{2} \bigintss_0^{2 \pi} \left[ \rho \left( \frac{\rho'}{2 \rho} - \frac{F_{\rho}}{D} \right)^2 + \rho^{-1} \frac{\omega^2}{4 \pi^2 D^2} + k \left(\rho - 1 \right) \right] \, d \theta \, ,
\end{equation}
where $k$ is the constant Lagrangian multiplier that fixes the normalisation constraint, i.e., $ 0 = \frac{\partial \mathcal{L}_1}{\partial k} = \int_0^{2 \pi} \rho \, d\theta - 1$. The Euler--Lagrange equation is calculated as follows:
\begin{equation}
\begin{split}
0 &= \left( \frac{d}{d \epsilon} \mathcal{L} \left[ \rho + \epsilon \eta, \rho' + \epsilon \eta' \right] \right)_{\epsilon=0} \\ 
&= \bigintss_0^{2 \pi} \left[ \eta \left( \left( \frac{\rho'}{2 \rho} - \frac{F_{\rho}}{D} \right)^2 + k - \frac{\rho^{-2} \omega^2}{4 \pi^2 D^2} \right) + 2 \rho \left( \frac{\rho'}{2 \rho} - \frac{F_{\rho}}{D} \right)\left( \frac{\eta'}{2 \rho} - \frac{2 \rho' \eta}{(2 \rho)^2} - \frac{2 c}{D} \int_0^{2 \pi} \sin(\theta_1 - \theta) \eta(\theta_1) \, d\theta_1 \right) \right] d \theta \, , \\
\end{split}
\end{equation}
and is valid for $\eta$-variations such that $\int_0^{2 \pi} \eta \, d\theta = 0$. After integrating by parts the term with $\eta'$, swapping integrals over the last term, and recognising that the resulting expression must hold for all $\eta$-variations, we obtain the following simplified Riccati-like expression for the Euler--Lagrange equation:
\begin{equation}
\label{eq:ELCurrLD}
0 =  \left( \frac{\rho'}{2 \rho} - \frac{F_{\rho}}{D} \right)^2 + k - \frac{\rho^{-2} \omega^2}{4 \pi^2 D^2} - \frac{d}{d \theta} \left( \frac{\rho'}{2 \rho} - \frac{F_{\rho}}{D} \right) - \frac{\rho'}{\rho} \left( \frac{\rho'}{2 \rho} - \frac{F_{\rho}}{D} \right) - \frac{4 c}{D} \int_0^{2 \pi} \rho \left( \frac{\rho'}{2 \rho} - \frac{F_{\rho}}{D} \right) \sin(\theta - \theta_1) \, d\theta_1 \, ,
\end{equation}
whose solution, namely $\rho^*$, is the minimiser of $I^{\text{(SID)}}_{\text{2.5}}$ for any fixed fluctuation $\omega$ of the current.

Evidently, the uniform distribution $\rho^*(\theta) = \rho_{\text{inv}}(\theta) = (2 \pi)^{-1}$ is always solution of \eqref{eq:ELCurrLD}. This means that whenever $\rho_{\text{inv}}(\theta) = (2 \pi)^{-1}$ is stable, i.e., in the self-repelling region, fluctuations of the SID current far from the typical value ($\omega \neq J_{\rho_{\text{inv}}}=0$) are generated by a process that preserves the uniform stationary distribution. Hence, such a process will simply be a drifted Brownian motion over the ring.

On the opposite, the localised distribution $\rho^*(\theta) = \rho_{\text{inv}}(\theta) = (\mathcal{Z})^{-1} e^{\frac{4 c}{D} \left( \cos (\theta) \alpha_1 + \sin (\theta) \alpha_2 \right)}$ is only solution for $\omega = 0$. Therefore, in the self-attracting region, fluctuations $\omega \neq 0$ are generated by a process whose stationary distribution $\rho^*$ is not given by $\rho_{\text{inv}}$. In the following we show how to derive the minimiser $\rho^*$ solution of \eqref{eq:ELCurrLD}.

\subsubsection{Perturbation theory around $\rho_{\text{inv}}$}

An exact analytic solution of \eqref{eq:ELCurrLD} is hard to find. For this reason, we consider the perturbative analytic approximation for small currents $\omega$ that follows:
\begin{equation}
\label{eq:DensitySmalljExpansion}
\rho = \rho_{\text{inv}} + \omega \tilde{\rho} + o(\omega) \ ,
\end{equation}
with the normalisation condition given by $\int_0^{2 \pi} \rho_{\text{inv}} \, d\theta = 1$ and $\int_0^{2 \pi} \tilde{\rho} \, d\theta = 0$. We do not know a priori whether a perturbative approach will work in general, nor whether the current should appear linearly in \eqref{eq:DensitySmalljExpansion}. In the worst case scenario, this approach will provide us with a weak(er) upper bound for the current rate function.

Replacing \eqref{eq:DensitySmalljExpansion} in \eqref{eq:ELCurrLD} and keeping only the first order contribution in $\omega$ (the zeroth one is $0$) we get
\begin{align}
\label{eq:1stOrderEl}
\begin{split}
0 &= \frac{\rho'_{\text{inv}} \tilde{\rho}'}{2} + \frac{2 \rho_{\text{inv}} \tilde{\rho} F_{\rho_{\text{inv}}}^2 + \rho_{\text{inv}}^2 F_{\rho_{\text{inv}}} F_{\tilde{\rho}}}{D^2} + k \rho_{\text{inv}} \tilde{\rho} - \frac{\tilde{\rho} \rho''_{\text{inv}} + \tilde{\rho}'' \rho_{\text{inv}}}{2} + \frac{2 \rho_{\text{inv}} \tilde{\rho} F'_{\rho_{\text{inv}}} + \rho_{\text{inv}}^2 F'_{\tilde{\rho}}}{D} + \\
&\hspace{5cm} - \frac{4 c}{D} \rho_{\text{inv}}^2 \int_0^{2 \pi} \left[ \sin(\theta - \theta_1) \rho_{\text{inv}}(\theta_1) \left( \frac{\tilde{\rho}'}{2 \rho_{\text{inv}}} - \frac{\rho'_{\text{inv}} \tilde{\rho}}{2 \rho_{\text{inv}}^2} - \frac{F_{\tilde{\rho}}}{D} \right)(\theta_1) \right] \, d \theta_1 \, ,
\end{split}
\end{align}
which is still a complicated expression to deal with. \newtext{Given $\rho_{\text{inv}}(\theta) = (\mathcal{Z})^{-1} e^{\frac{4 c}{D} \left( \cos (\theta) \alpha_1 + \sin (\theta) \alpha_2 \right)}$, $F_{\rho_{\text{inv}}}$ and $F_{\tilde{\rho}}$ defined as in \eqref{eq:SIDStatFokkerPlanckAsymptDrift}}, we further consider the following \newtext{exact} relations:
\begin{align}
\rho'_{\text{inv}}(\theta) &= \rho_{\text{inv}}(\theta) \frac{4 c}{D} (\alpha_2 \cos \theta - \alpha_1 \sin \theta) \\
\rho''_{\text{inv}}(\theta) &= \rho_{\text{inv}}(\theta) \frac{16 c^2}{D^2} (\alpha_2 \cos \theta - \alpha_1 \sin \theta)^2 + \rho_{\text{inv}}(\theta) \frac{4 c}{D} (-\alpha_2 \sin \theta - \alpha_1 \cos \theta) \\
F_{\rho_{\text{inv}}}(\theta) &= - \alpha_1 \sin \theta + \alpha_2 \cos \theta \\
F'_{\rho_{\text{inv}}}(\theta) &= 2 c (- \alpha_1 \cos \theta - \alpha_2 \sin \theta) \\
F_{\tilde{\rho}}(\theta) &= 2c (\bar{\alpha}_2 \cos \theta - \bar{\alpha}_1 \sin \theta) \\
F'_{\tilde{\rho}}(\theta) &= 2c (- \bar{\alpha}_2 \sin \theta - \bar{\alpha}_1 \cos \theta) \, ,
\end{align}
along with the definitions
\begin{align}
\label{eq:ab1}
\bar{\alpha}_1 &= \int_0^{2 \pi} \tilde{\rho}(\theta) \cos \theta \, d \theta \\
\label{eq:ab2}
\bar{\alpha}_2 &= \int_0^{2 \pi} \tilde{\rho}(\theta) \sin \theta \, d \theta \\
\label{eq:ab3}
\bar{\alpha}_3 &= \int_0^{2 \pi} \tilde{\rho}(\theta) \cos^2 \theta \, d \theta \\
\label{eq:ab4}
\bar{\alpha}_4 &= \int_0^{2 \pi} \tilde{\rho}(\theta) \cos \theta \sin \theta \, d \theta \\
\label{eq:ab5}
\bar{\alpha}_5 &= \int_0^{2 \pi} \tilde{\rho}(\theta) \sin^2 \theta \, d \theta \\
\label{eq:a3}
\alpha_3 &= \int_0^{2 \pi} \rho_{\text{inv}}(\theta) \cos^2 \theta \, d \theta \\
\label{eq:a4}
\alpha_4 &= \int_0^{2 \pi} \rho_{\text{inv}}(\theta) \cos \theta \sin \theta \, d \theta \\
\label{eq:a5}
\alpha_5 &= \int_0^{2 \pi} \rho_{\text{inv}}(\theta) \sin^2 \theta \, d \theta \, ,
\end{align}
and obtain, after some simplifications, the following equation for the perturbed density $\tilde{\rho}$:

\begin{equation}
\label{eq:Rho1Eq}
\begin{split}
0 &= - \tilde{\rho}''(\theta) + \frac{4 c}{D} \tilde{\rho}'(\theta) (\alpha_2 \cos \theta - \alpha_1 \sin \theta) - \frac{4 c}{D} \tilde{\rho}(\theta) (\alpha_1 \cos \theta + \alpha_2 \sin \theta) + g(\theta) \, , 
\end{split}
\end{equation}
with
\begin{equation}
\label{eq:ExplicitTermg}
\begin{split}
g(\theta) &= \frac{4 c}{D} \rho_{\text{inv}}(\theta) \Bigg[ \frac{2 c}{D} (\alpha_2 \cos \theta - \alpha_1 \sin \theta )(\bar{\alpha}_2 \cos \theta - \bar{\alpha}_1 \sin \theta) - 2 (\bar{\alpha}_2 \sin \theta + \bar{\alpha}_1 \cos \theta) + \\
&\hspace{0.2cm} + \frac{4 c}{D} \left( \alpha_2 (\bar{\alpha}_3 \sin \theta - \bar{\alpha}_4 \cos \theta ) + \alpha_1 (\bar{\alpha}_5 \cos \theta - \bar{\alpha}_4 \sin \theta) + \bar{\alpha}_2 (\alpha_3 \sin \theta - \alpha_4 \cos \theta ) + \bar{\alpha}_1 (\alpha_5 \cos \theta - \alpha_4 \sin \theta) \right) \Bigg] \ .
\end{split} 
\end{equation}

\newtext{The differential equation \eqref{eq:Rho1Eq} will be solved by imposing the boundary condition $\tilde{\rho}(0) = \tilde{\rho}(2\pi)$ and the global condition $\int_0^{2\pi} \tilde{\rho} = 0$, both of which naturally arise from the requirement that $\tilde{\rho}$ is a perturbation of $\rho_{\text{inv}}$, as in \eqref{eq:DensitySmalljExpansion}.}

The homogeneous solution of \eqref{eq:Rho1Eq} is
\begin{equation}
\label{eq:GeneralRho1}
\tilde{\rho}^{\text{G}} (\theta) = c_1 y_1 (\theta) + c_0 y_0 (\theta) \ ,
\end{equation}
with
\begin{align}
\label{eq:y0}
y_0(\theta) &= \mathcal{Z} \rho_{\text{inv}} (\theta) e^{-\frac{4 c}{D} \alpha_1} \\
\label{eq:y1} 
\begin{split}
y_1(\theta) &= \mathcal{Z} \rho_{\text{inv}} (\theta)  \left( \int_0^\theta e^{- \frac{4 c}{D} ( \alpha_1 \cos \theta_1 + \alpha_2 \sin \theta_1)} \, d \theta_1 \right) \, ,
\end{split}
\end{align}
where we remind the reader that $\mathcal{Z} = \int_0^{2 \pi} e^{\frac{4 c}{D} \left( \cos \theta \alpha_1 + \sin \theta \alpha_2 \right)} \, d\theta$.
Using the Wronskian $W(y_1,y_0) = y_1 y'_0 - y_0 y'_1 = - y_0$, we can find a particular solution to the complete heterogeneous problem in \eqref{eq:Rho1Eq}, which reads
\begin{equation}
\label{eq:ParticularRho1}
\begin{split}
\tilde{\rho}^{\text{P}} (\theta) =  y_1(\theta) \int_0^{2 \pi}  g(\theta) \, d \theta - y_0 (\theta) \int_0^{2 \pi} \frac{y_1(\theta) g(\theta)}{y_0(\theta)} \, d \theta \, .
\end{split}
\end{equation}

The final (formal) solution of \eqref{eq:Rho1Eq} is given by
\begin{equation}
\label{eq:FinalSolution}
\begin{split}
\tilde{\rho}(\theta) &= \tilde{\rho}^{\text{G}}(\theta) + \tilde{\rho}^{\text{P}}(\theta) \\
&= \rho_{\text{inv}}(\theta) \left( \tilde{c}_0 + c_1 \int_0^{\theta}  \rho_{\text{inv}}^{-1}(\theta_1) \, d \theta_1 + \int_0^{\theta}  \rho_{\text{inv}}^{-1}(\theta_1) \, d \theta_1 \int g(\theta) \, d \theta - \int \left[ g(\theta) \int_0^{\theta} \rho_{\text{inv}}^{-1}(\theta_1) \, d \theta_1 \right] \, d \theta \right) \ ,
\end{split}
\end{equation}
with $\tilde{\rho}^{\text{G}}$ given in \eqref{eq:GeneralRho1} and $\tilde{\rho}^{\text{P}}$ in \eqref{eq:ParticularRho1} (and $\tilde{c}_0 = \mathcal{Z} e^{- \frac{4 c}{D} \alpha_1} c_0$ is just another constant). Eq.\ \eqref{eq:FinalSolution} along with \eqref{eq:SIDStatFokkerPlanckAlpha1}, \eqref{eq:SIDStatFokkerPlanckAlpha2}, and \eqref{eq:ab1}--\eqref{eq:a5} form a closed system of equations that can be solved to explicitely determine \eqref{eq:FinalSolution}. Additionally, by imposing the condition $\tilde{\rho}(0) = \tilde{\rho}(2 \pi)$ and considering the notation $G(\theta) = \int g(\theta) \, d\theta$ and $\bar{G}(\theta) = \mathcal{Z}^{-1} \int \left[ g(\theta) \int_0^\theta \rho^{-1}_{\text{inv}}(\theta_1) d \theta_1 \right] \, d\theta$ we obtain the constant
\begin{equation}
\label{eq:c1Constant}
c_1 =\mathcal{Z}  e^{-\frac{4 c}{D} \alpha_1}  \left( \frac{\bar{G}(2 \pi) - \bar{G}(0)}{ \int_0^{2 \pi} \rho^{-1}_{\text{inv}}(\theta) \, d \theta} \right) - G(2 \pi) \, .
\end{equation}
Eventually, the constant $c_0$ (or $\tilde{c}_0$) is fixed by imposing the normalisation constraint $\int_0^{2 \pi} \tilde{\rho}(\theta) \, d \theta = 0$. \newtext{As a final remark, we observe that due to the form of the solution \eqref{eq:FinalSolution}, the differential equation \eqref{eq:Rho1Eq} remains invariant under rescaling by any constant. Therefore, the specific value of the function in Fig.\ \ref{fig:pert} is not significant; what matters is the shape of the function and its interpretation, which will be discussed in the next subsection. In fact, we expect that $\tilde{\rho}$ becomes arbitrarily small as $\omega \rightarrow 0$, in line with Eq.\ \eqref{eq:DensitySmalljExpansion}.}

\newtext{
\begin{figure}
\centering
\includegraphics[width=\textwidth]{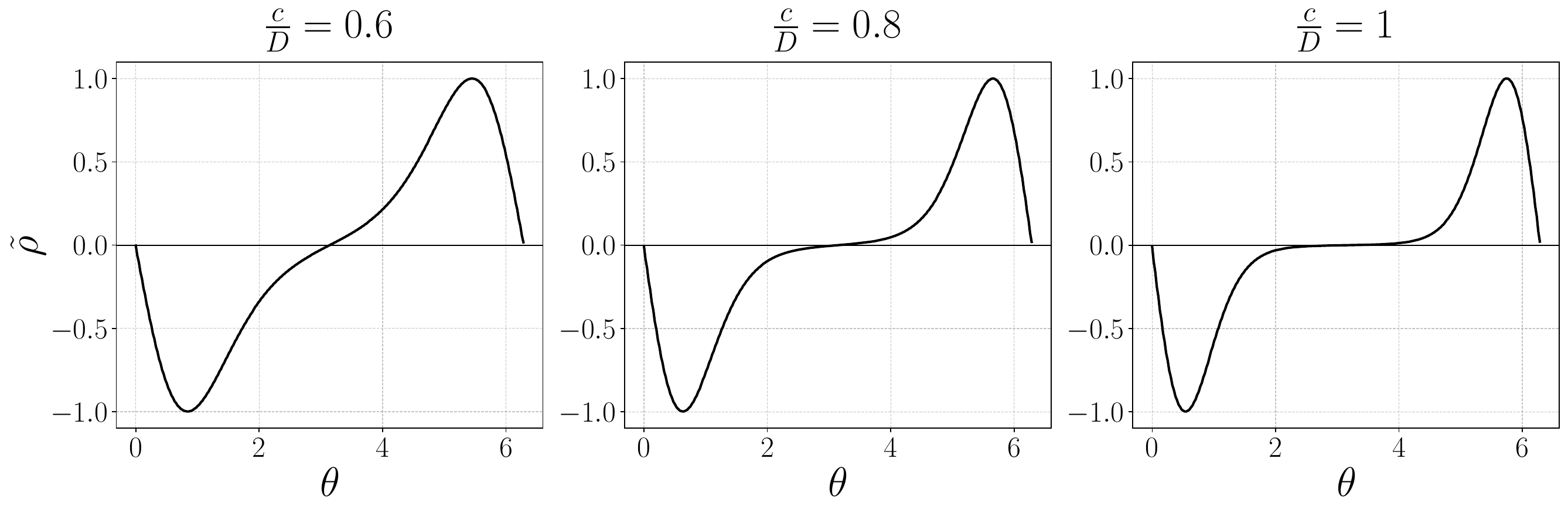}
\caption{Exemplary graphs of $\tilde{\rho}(\theta)$ for $\rho_{\text{inv}}(\theta) = (\mathcal{Z})^{-1} e^{\frac{4 c r}{D} \cos (\theta)}$---obtained setting $\alpha_1 = r$ solution of \eqref{eq:O2SymmetryPhiPi} and $\alpha_2=0$---localised where the SID starts, at increasing values of $c/D$. }
\label{fig:pert}
\end{figure}}

To summarise, the function $\tilde{\rho}$ in \eqref{eq:FinalSolution} is the small-current change to the stationary distribution $\rho_{\text{inv}}$ characterising the optimal solution to the minimisation problem \eqref{eq:InfimumRhoLevel2.5}. Therefore, according to the driven process theory~\cite{Chetrite2015} for a small current $\omega$, the density $\rho^* = \rho_{\text{inv}} + \omega \tilde{\rho}$ characterises the stationary distribution of the auxiliary process that typically generates the current $\omega$. Notice that, because of the first assumption made in \ref{subsubsec:SIDs}, i.e., the existence of an auxiliary dynamics which typically converges to a fluctuation of $\rho_t$ and $j_t$, and the general linear-in-$\omega$ form of the expansion \eqref{eq:DensitySmalljExpansion}, the inherently-Markovian dynamics of the auxiliary process just determined may not be the most likely dynamics that generate the current $\omega$ in an SID. Therefore, in this sense, the auxiliary process is only sub-optimal, as much as the rate function $I(\omega)$ in \eqref{eq:InfimumRhoLevel2.5} is expected to only be an upper bound of the real current rate function for the SID.

\subsubsection{Current rate function for small $\omega$}

As mentioned in the last paragraph of \ref{subsubsec:SIDs}, in the self-repelling region fluctuations of the SID current are generated by an auxiliary drifted Brownian dynamics over the ring whose stationary distribution is uniform. Therefore, we expect $\tilde{\rho} = 0$. To check this we plug $\rho_{\text{inv}} = (2 \pi)^{-1}$, along with $\alpha_1 = \alpha_2 = \alpha_4 = 0$ and $\alpha_3 = \alpha_5 = 1/2$ from \eqref{eq:SIDStatFokkerPlanckAlpha1}, \eqref{eq:SIDStatFokkerPlanckAlpha2} and \eqref{eq:a3}--\eqref{eq:a5}, in \eqref{eq:FinalSolution} and obtain 
\begin{equation}
\label{eq:RhoTildeErgodic}
\tilde{\rho}(\theta) = \left(\frac{4 c}{\pi D} - \frac{4 c^2}{\pi D^2} \right) \left( \bar{\alpha}_1 \cos \theta + \bar{\alpha}_2 \sin \theta \right) \, ,
\end{equation}
where $\bar{\alpha}_1$ and $\bar{\alpha}_2$ are the only two parameters left to be determined. Solving for them from \eqref{eq:ab1} and \eqref{eq:ab2} we get $\bar{\alpha}_1 = \bar{\alpha}_2 = 0$ and, as a consequence, $\tilde{\rho}=0$ as expected.

In the self-attracting region where a localised distribution $\rho_{\text{inv}}$ is reached, the form of $\tilde{\rho}$ will have to be determined by solving the system of equations given by \eqref{eq:FinalSolution} along with \eqref{eq:SIDStatFokkerPlanckAlpha1}, \eqref{eq:SIDStatFokkerPlanckAlpha2} and \eqref{eq:ab1}--\eqref{eq:a5}.
To build intuition, we plot exemplary graphs of $\tilde{\rho}(\theta)$ in Fig.\ \ref{fig:pert} for $\rho_{\text{inv}}(\theta) = (\mathcal{Z})^{-1} e^{\frac{4 c r}{D} \cos (\theta)}$ localised at the origin of the ring, where the SID starts, at increasing values of $c/D$. From the plots, we notice that the form of $\tilde{\rho}$ is always antisymmetric with respect to the origin with a positive part just to the left of the origin and a negative one to the right. This means that the final form of $\rho^* = \rho_{\text{inv}} + \omega \tilde{\rho}$, which generates the small current $\omega$ will still be highly localised at $0$, but with a small bulge on the left and a dip on the right. This form breaks the symmetry of $\rho_{\text{inv}}$ and favours the generation of a counterclockwise current $\omega$. The opposite behaviour, i.e., clockwise current, is generated similarly by considering the change $\tilde{\rho} \rightarrow - \tilde{\rho}$. Additionally, from Fig.\ \ref{fig:pert} we also notice that the higher the ratio $c/D$ (greater localisation) the thinner the peaks of $\tilde{\rho}$ and therefore the more localised the bulge and the dip in $\rho^*$.

Now that we have a perturbative form for the optimal $\rho^*$ solution of \eqref{eq:InfimumRhoLevel2.5}, we can use this to determine the shape of the rate function $I(\omega)$ for small currents $\omega$. To do so, we take the ansatz \eqref{eq:DensitySmalljExpansion} (plus a `bookkeeping' term of $O(\omega^2)$ that will anyway disappear), plug it into $I^{\text{(SID)}}_{\text{2.5}}[\rho,\omega]$ in \eqref{eq:Level2.5SIDRingFinal} and keep terms up to $O(\omega^2)$ to eventually get 
\begin{align}
I_{\text{c}}(\omega) &= I^{\text{(SID)}}_{\text{2.5}}[\rho_{\text{inv}} + \omega \tilde{\rho} + \frac{\omega^2}{2} \tilde{\tilde{\rho}} + o(\omega^2)] \\
\label{eq:RateFunctionCurrentExp}
&= \frac{\omega^2}{2} D \int_0^{2 \pi} \left[ \rho_{\text{inv}}(\theta) \left( \frac{1}{2} \frac{d}{d\theta}\left( \frac{\tilde{\rho}(\theta)}{\rho_{\text{inv}}(\theta)} \right) - \frac{F_{\tilde{\rho}}(\theta)}{D} \right)^2 + \frac{\rho_{\text{inv}}^{-1}(\theta)}{4 \pi^2 D^2} \right] \, d \theta + o(\omega^2) \, .
\end{align}
In doing so, we realise that the form of the rate function $I_{\text{c}}(\omega)$ should explicitly depend on the parameters $\alpha_1$ and $\alpha_2$, which fully determine $\rho_{\text{inv}}$ (along with other parameters) around which we are expanding. However, as we explain later, this dependence does not change the form of the rate function for small currents, and therefore, we do not keep track of it. We remark that the factor
\begin{equation}
\label{eq:AsymptoticVariance}
\left( D \int_0^{2 \pi} \left[ \rho_{\text{inv}} \left( \frac{1}{2} \frac{d}{d\theta} \left( \frac{\tilde{\rho}}{\rho_{\text{inv}}} \right) - \frac{F_{\tilde{\rho}}}{D} \right)^2 + \frac{\rho_{\text{inv}}^{-1}}{4 \pi^2 D^2} \right] \right)^{-1} \eqqcolon \sigma^2_{\text{c}}[\tilde{\rho}] \, ,
\end{equation}
which multiplies $\omega^2/2$ in \eqref{eq:RateFunctionCurrentExp}, is a lower bound on the so-called asymptotic variance, i.e,
\begin{equation}
\label{eq:AsympVarGenDef}
\sigma^2 \coloneqq \lim_{t \rightarrow \infty} t \mathbb{E} \left[ (\Omega_t - J_{\rho_{\text{inv}}})^2 \right] \, ,
\end{equation}
of the current of the SID process converging to $\rho_{\text{inv}}$. Evidently, from the quadratic form of $I_c(\omega)$, fluctuations $\omega$ are Gaussian, with a width determined by \eqref{eq:AsymptoticVariance}, around $J_{\rho_{\text{inv}}} = 0$.

\section{Results}
\label{sec:results}

In this Section, we collect all the results obtained studying the fluctuations of the current $\Omega_t$ in \eqref{eq:CurrObs} using the methods introduced in Sec.\ \ref{sec:methods}.

\subsection{Asymptotic variance}
\label{subsec:AsymptoticVariance}

\begin{figure}[ht]
    \centering
    \begin{subfigure}[b]{0.49\textwidth}
        \centering
        \includegraphics[width=\textwidth]{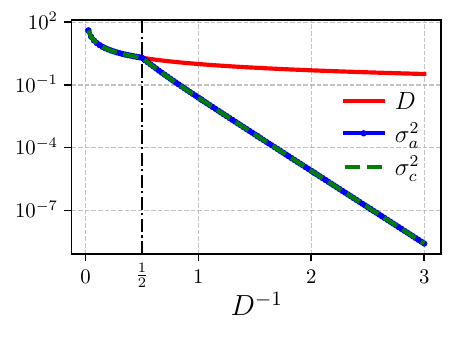}
    \end{subfigure}
    \hfill
    \begin{subfigure}[b]{0.49\textwidth}
        \centering
        \includegraphics[width=\textwidth]{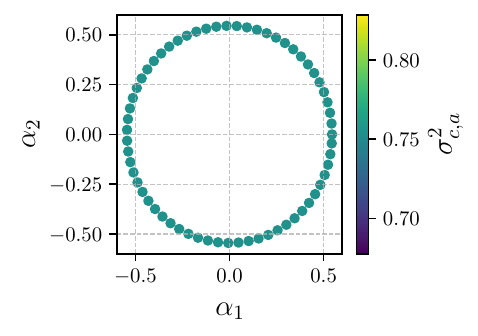}
    \end{subfigure}
\caption{\newtext{On the left,} lower bound $\sigma^2_{\text{c}}$ on the asymptotic variance of the SID in \eqref{eq:AsymptoticVariance} and asymptotic variance $\sigma^2_{\text{a}}$ of the adiabatic process in \eqref{eq:SIDAdiabatic} compared with the diffusion coefficient $D$ of a pure Brownian motion. Given $c/D$, $\sigma^2_{\text{c}}$ and $\sigma^2_{\text{a}}$ are both constant with varying $\alpha_1$ and $\alpha_2$ such that $\alpha_1^2 + \alpha_2^2 = r^2$ with $r$ solution of \eqref{eq:O2SymmetryPhiPi}. [For the plot we chose $\alpha_1 = r$ and $\alpha_2=0$.] \newtext{On the right, an exemplary plot demonstrates the invariance of $\sigma^2_{\text{c}}$ and $\sigma^2_{\text{a}}$ with varying $\alpha_1$ and $\alpha_2$ at a fixed $c/D = 0.6$.}}
\label{fig:asymptvar}
\end{figure}

We discuss the behaviour of the asymptotic variance in \eqref{eq:AsymptoticVariance} in both the self-repelling ($c/D \leq 1/2$) and self-attracting ($c/D > 1/2$) region. In the self-repelling region, $\rho_{\text{inv}}(\theta) = (2 \pi)^{-1}$ and $\tilde{\rho} = 0$, and therefore
\begin{equation}
\label{eq:AsymptoticVarianceErgodic}
\sigma^2_{\text{c}}[0] = D \, ,
\end{equation}
proving once again that the auxiliary process determining small-current fluctuations is a simple drifted Brownian motion over the ring. In the self-attracting region, for generic values of $c$ and $D$, $\sigma^2_{\text{c}}$ can only be evalueated numerically. Remarkably, although $\sigma^2_{\text{c}}[\tilde{\rho}]$ has a `hidden' dependence on $\rho_{\text{inv}}$ through the parameters $\alpha_1$ and $\alpha_2$ that fully determine $\rho_{\text{inv}}$, we can show numerically that $\sigma^2_{\text{c}}$ does not change with varying $\alpha_1$ and $\alpha_2$ given the relation $\alpha_1^2 + \alpha_2^2 = r^2$ with $r$ solution of \eqref{eq:O2SymmetryPhiPi}. \newtext{[We show this for an exemplary value of $c/D = 0.6$ in Fig.\ \ref{fig:asymptvar} (right), but the same result holds for any other values of $c/D$]}. In other words, the $O(2)$ symmetry of $\rho_{\text{inv}}$ is inherited by the rate function and, in particular, by the asymptotic variance.

In Fig.\ \ref{fig:asymptvar} \newtext{(left)}, we compare the diffusion coefficient of a pure Brownian motion $dB_t = \sqrt{D} dW_t$ with the lower bound $\sigma^2_{\text{c}}$ we have on the asymptotic variance of the SID in \eqref{eq:AsymptoticVariance} and with the asymptotic variance $\sigma^2_{\text{a}}$ of the adiabatic process in \eqref{eq:SIDAdiabatic}. The latter is calculated as $\sigma^2_{\text{a}} = \lambda_a''(k)|_{k=0}$, where $\lambda_a$ is given in \eqref{eq:SCGF}. It shows numerically the same $O(2)$ symmetry of $\sigma^2_{\text{c}}$ and is therefore invariant as well with respect to the choice of $\alpha_1$ and $\alpha_2$ \newtext{(see Fig.\ \ref{fig:asymptvar} (right))}. Evidently, for $c/D < 1/2$ we have \eqref{eq:AsymptoticVarianceErgodic} and the three curves fully overlap. On the other hand, for $c/D > 1/2$, $\sigma^2_c$ and $\sigma^2_{\text{a}}$ start decreasing exponentially fast, deviating from the pure diffusive behaviour. A sudden transition in the behaviour of $\sigma^2_c$ and $\sigma^2_{\text{a}}$ arises at $c/D = 1/2$, which we believe is consequence of the delocalisation-localisation phase transition discussed in \ref{subsubsec:PhaseDiagram}. 

The current state of our research does not include an analytical proof to support the claim of the transition at the level of the asymptotic variance \newtext{and neither a proof that shows that $\sigma_a^2 = \sigma_c^2$ within the $O(2)$ symmetry and for all $c/D$}. However, the numerical results are clear. Furthermore, the overall description aligns with the transition observed in~\cite{Delgadino2021} (see Sec.\ 1.10 and Eq.\ 1.34) regarding the transport coefficient of a noisy mean-field Kuramoto model. This model's associated Hamiltonian is the same as that of the $XY$ model discussed in the last paragraph of \ref{subsubsec:PhaseDiagram} and in Appendix \ref{appendix:XY}). 


Additionally, the fact that the lower bound on the asymptotic variance $\sigma^2_{\text{c}}$ of the SID always overlap with the asymptotic variance of the adiabatic process $\sigma^2_{\text{a}}$ means that for small current fluctuations the adiabatic process and the SID behave similarly at long times. A more detailed comparison of these fluctuations follows.

\subsection{Comparison with Monte-Carlo simulations}
\label{subsec:monte}

In the self-repelling scenario, the long-time behaviour of the current observable $\Omega_t$ in \eqref{eq:CurrObs} associated to an SID process is equivalent to considering the ensemble average. In general, in a self-attracting scenario, this is no longer true. By assuming an ensemble point of view, i.e., considering infinitely many realisations of the SID process in \eqref{eq:SIDEmp}, the distribution of the current $P_e(\Omega_t = \omega)$ can be written as
\begin{equation}
\label{eq:EnsembleCurrent}
P_e(\Omega_t = \omega) = \int_{\alpha_1^2 + \alpha_2^2 = r^2} P_e(\Omega_t = \omega| a_{t,1} = \alpha_1, a_{t,2} = \alpha_2 ) P(a_{t,1} = \alpha_1, a_{t,2} = \alpha_2 ) \, d\alpha_1 d\alpha_2 ,
\end{equation}
where we have introduced new random variables $a_{t,1}$ and $a_{t,2}$, whose values depend on the entire realisation of the noise up to time $t$ in \eqref{eq:SIDEmp}, along with their joint distribution, and made use of the law of conditioning. In general, a different convergence of the empirical measure of the SID, fully determined by the values $\alpha_1$ and $\alpha_2$, could lead to different fluctuations in the current. 


On the other hand, our analytical and numerical results for the long-time limit of $P(\Omega_t = \omega)$ show no dependence on $\alpha_1$ and $\alpha_2$, i.e., the large deviation form
\begin{equation}
\label{eq:LargeDeviationCurrent}
P(\Omega_t = \omega) \approx e^{- t I(\omega)} \, ,
\end{equation}
holds, and the rate functions $I \equiv I_{\text{a}}$ or $I \equiv I_{\text{c}}$, actually only the small-current limit of the latter, are invariant with respect to a change of $\alpha_1$ and $\alpha_2$ as mentioned above. We therefore expect that by running long simulations 
\begin{equation}
\label{eq:EnsembleCurrentLong}
P_e(\Omega_t = \omega) \approx P(\Omega_t = \omega) \approx e^{- t I_{\text{a}/\text{c}}(\omega)} \, ,
\end{equation}
which means that the current observable behaves ergodically.

\begin{figure}
\includegraphics[width=0.85\textwidth]{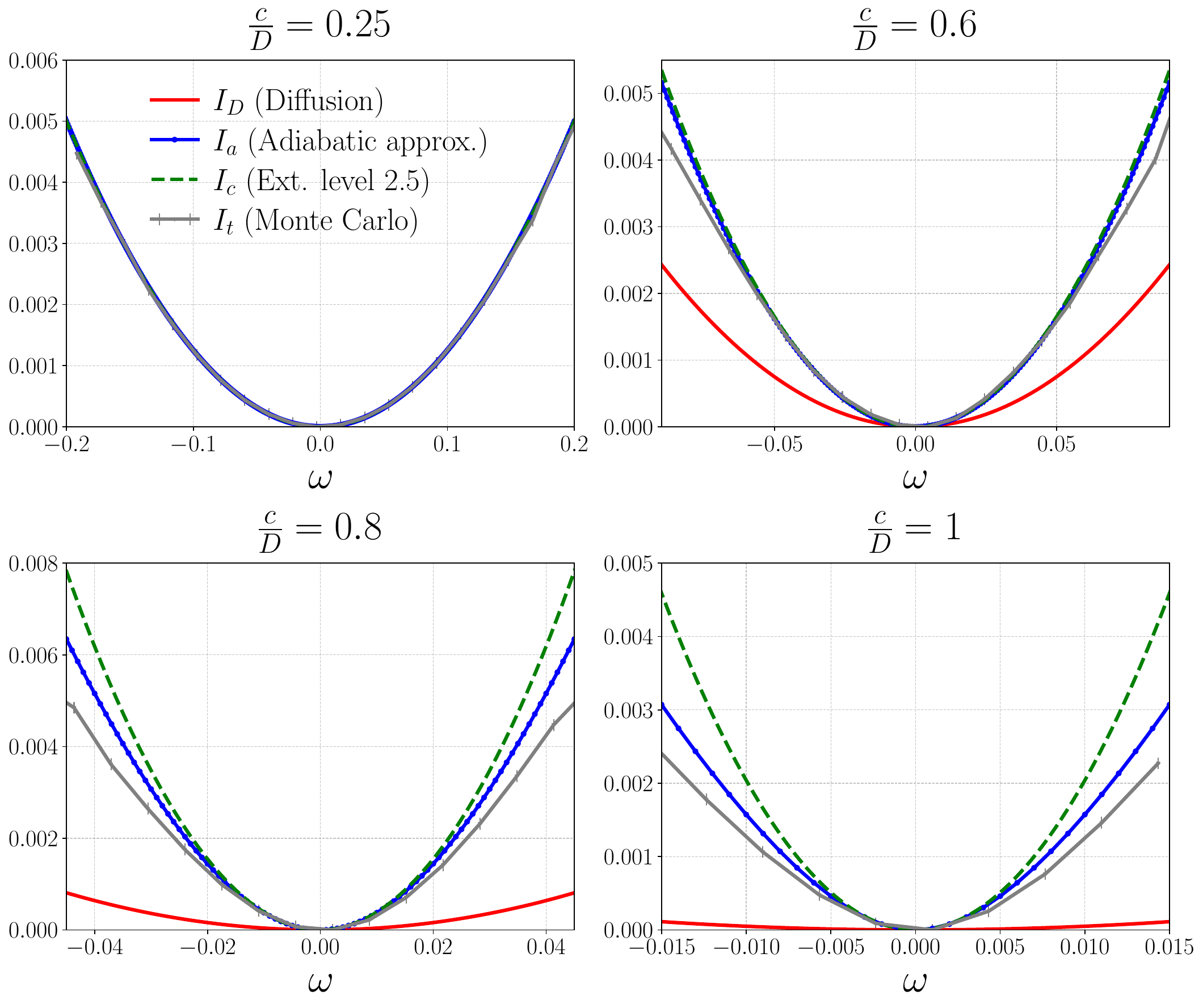}
\caption{Monte Carlo simulations, plotted as $I_t(\omega)$ in \eqref{eq:RateFunctionMonteCarlo}, were obtained by running \newtext{$10^6$ times the SID process in \eqref{eq:SIDEmp} for $t=2000$} with $dt=0.025$, c=$1$, and $D=4, 1.67, 1.25, 1$ respectively for the four plots from top left to bottom right. \newtext{Error bars are computed by constructing normal `square-root' error bars for the histogram and by transforming them according to \eqref{eq:RateFunctionMonteCarlo}}. These simulations are compared with the current rate functions: $I_a$ in \eqref{eq:LegendreFenchel} for the adiabatic process, $I_c$ derived from extended level 2.5 large deviation theory in the small-current limit of \eqref{eq:RateFunctionCurrentExp}, and $I_D(\omega)$ of a pure Brownian motion over the ring.}
\label{fig:monte}
\end{figure}

To check this statement and our analytical and numerical results above, we run Monte Carlo simulations for long times $t$ for different values of $c/D$. \newtext{Monte Carlo simulations are obtained by running the Euler--Maruyama discretised version of the SID in \eqref{eq:SIDEmp}. For details, we refer the reader to Appendix \ref{appendix:MonteCarlo} and to the caption of Fig.\ \ref{fig:monte}.} For each simulation, starting from the histogram of currents $f_t(\omega)$, which is expected to converge (at long times) to $P_e(\Omega_t = \omega)$ in \eqref{eq:EnsembleCurrent}, the function 
\begin{equation}
\label{eq:RateFunctionMonteCarlo}
I_t(\omega) = - \frac{1}{t} \ln f_t(\omega) \, 
\end{equation}
should converge to the true rate function of the SID process. In Fig.\ \ref{fig:monte}, we compare this function with the current rate functions $I_a$ in \eqref{eq:LegendreFenchel} associated to the adiabatic process, $I_c$ derived from the contraction over the extended level 2.5 large deviation theory in the small-current limit of \eqref{eq:RateFunctionCurrentExp}, and $I_D(\omega) = \omega^2/(2D)$ of a pure Brownian motion over the ring. 

From the plots, we notice that in the self-repelling phase ($c/D \leq 1/2$), all curves and simulation data points perfectly overlap: long-time fluctuations of the current are simply generated by pure Brownian motion over the ring. In the self-attracting phase ($c/D > 1/2$), we investigate three scenarios with increasing localisation: $c/D = 0.6$, $c/D = 0.8$, and $c/D = 1$. Recall that the greater the localisation the sharper $\rho_{\text{inv}}$ around its typical value. In all cases, $I_D$ greatly underestimates $I_t$ in \eqref{eq:RateFunctionMonteCarlo} providing a lower-bound that becomes weaker the greater the localisation. For weak localisation, both $I_a$ and $I_c$ well estimate simulations. For strong localization, \newtext{both $I_c$ and $I_a$ serve as upper bounds for $I_t$, with $I_a$ consistently providing a tighter bound than $I_c$. We anticipated $I_c$ to be an upper bound due to the way it was derived. However, the adiabatic approximation $I_a$ is much less controlled, and we can only assert that it is an upper bound a posteriori, by inspecting the plots in Fig.\ \ref{fig:monte}.

Although one might intuitively expect $I_c$ to provide a better approximation than $I_a$---since it allows for deviations from $\rho_{\text{inv}}$---it is important to note that the derivation of $I_c$ is based on the extended level 2.5 large deviations, which already serves as an upper bound on the true large deviation function for both the density and current of the SID.}

For small current fluctuations, the rate functions $I_a$ and $I_c$ are equivalent, as indicated by the asymptotic variances in Fig.\ \ref{fig:asymptvar}. \newtext{In conclusion, both $I_a$ and $I_c$ are upper bounds on $I_t$, but in the localisation regimes analysed, the adiabatic approximation $I_a$ provides a tighter bound than $I_c$, more accurately capturing the behaviour of current fluctuations. In contrast, both $I_c$ and $I_a$ are equally tight in the self-repelling region.}

\section{Conclusion}
\label{sec:conclusion}

In this study, we have advanced the understanding of fluctuations in non-Markov processes by examining \newtext{for the first time} the average speed or current of a self-interacting diffusion on a ring. Our analysis revealed a delocalisation-localisation phase transition characterised by a shift from a self-repelling to a self-attracting regime. We employed two methods to analyse the current fluctuations: the adiabatic approximation and an original extension of level 2.5 large deviation theory along with perturbation theory.

Monte Carlo simulations confirmed our analytical predictions, showing that in the self-repelling phase ($c/D < 1/2$), the fluctuations of the current are well-described by pure Brownian motion over the ring. In the self-attracting phase ($c/D > 1/2$), increasing localisation sharpened the stationary distribution $\rho_{\text{inv}}$ around its typical value, influencing the fluctuation behaviour. Our results indicated that while $I_D$, the current rate function of a pure Brownian motion, can only be accounted as a weak lower bound of simulations $I_{\text{t}}$, the adiabatic approximation $I_{\text{a}}$ and $I_{\text{c}}$ derived from the extended level 2.5 along with perturbation theory \newtext{serve as upper bounds}, especially tight for weakly localised regimes. Finally, both the methods employed accurately estimated the asymptotic variance and suggested a phase transition at the onset of the localised regime.

Several open questions remain for future research. One key area of interest is determining whether rare events, beyond small-to-moderate current fluctuations, can be effectively captured by the adiabatic approximation, accurately estimated using the extended level 2.5 large deviation theory, and investigated in the large deviation regime through new numerical sampling methods tailored for non-Markov processes. Initial and transient noise paths could significantly influence the memory term, potentially leading to rare events that are not easily detected by the current framework. Additionally, gaining a deeper understanding of the non-stationary behaviour in the white region of the phase diagram, where the stationary Fokker--Planck equation fails, is crucial. Exploring how different potential forms and interaction strengths affect the fluctuation behaviour of SIDs could also provide valuable insights.

\newtext{We believe the methods introduced in this paper could be applied to a broader class of self-interacting processes, as evidenced by the recent derivation of a similar upper bound for discrete-time self-interacting chains in \cite{Budhiraja2023}. The particle on a ring, studied here, serves as a concrete example where calculations can be carried out more explicitly. In general, there is a strong need to better understand how fluctuations arise in non-Markovian systems, whether these systems follow a ‘pure’ non-Markovian dynamics or exhibit non-Markovian behaviour due to coarse-graining, as often occurs in experimental settings. Extending these methods to other systems could help uncover universal properties of self-interacting diffusions.}

Lastly, alongside theoretical work, there is considerable scope for applied research, particularly in understanding fluctuations of relevant observables in reinforcement mechanisms used in autochemotactic agents~\cite{DAlessandro2021}, swarm intelligence~\cite{Kennedy2001}, and machine learning algorithms~\cite{Sutton1998}.









\section*{Acknowledgments}

The author is deeply grateful to R.\ Chetrite for introducing him to the topic and for the initial discussions and insights that fueled this work. The author also expresses gratitude to R.\ Harris, Sarah A.\ M.\ Loos, K.\ S.\ Olsen, G.\ Pavliotis, H.\ Touchette and P.\ Vivo. for their engaging and enlightening discussions. This work was partially supported by the Swedish Research Council Grant No.\ 638-2013-9243.

\appendix

\newtext{
\section{Fokker--Planck equation for $(\theta_t, \rho_t)$}
\label{appendix:FokkerPlanck}

The stochastic process \eqref{eq:SIDEmpPot} is not Markovian because its transition probability function depends on the empirical occupation measure $\rho_t$. However, it is enough to extend the state space to include $\rho_t$ to study the joint stochastic process $(\theta_t,\rho_t)_{t \geq 0}$ as Markov.

Generally speaking, the dynamics of the joint stochastic process is given by
\begin{equation}
\label{eq:JointDynamicsGeneral}
\begin{cases}
	d\rho_t(\theta) &= \frac{dt}{t} \left( \delta_{\theta_t,\theta} - \rho_t(\theta) \right) \\
    d\theta_t &= \sigma dW_t + F_{\rho_t}(\theta_t) \, ,
\end{cases}
\end{equation}
and due to the nature of $\rho_t$, the dynamics lives in an infinite-dimensional state space. However, due to the peculiar form of the drift $F_{\rho_t}(\theta_t)$ in \eqref{eq:DriftRhoRing}, we can  greatly simplify \eqref{eq:JointDynamicsGeneral}. We start by using the addition formula for the $\sin$ function, which yields
\begin{equation}
\label{eq:AdditionSine}
\begin{split}
F_{\rho_t}(\theta_t) &= - 2 c \left( \int_0^{2\pi} \sin (\theta - \theta_t + \phi) \rho_t(\theta) \, d\theta \right) \\
&= - 2 c \left( \alpha_{2,t}^\phi \cos \theta_t - \alpha_{1,t}^\phi \sin \theta_t \right) \, ,
\end{split}
\end{equation}
where
\begin{equation}
\label{eq:DefinitionAlphat}
\begin{cases}
\alpha_{1,t}^\phi &= \int_0^{2\pi} \cos(\theta + \phi) \rho_t(\theta) \, d\theta \\
\alpha_{2,t}^\phi &= \int_0^{2\pi} \sin(\theta + \phi) \rho_t(\theta) \, d\theta \, .
\end{cases}
\end{equation}
Hence, the initial dependence on the distribution $\rho_t$ is simplified to a 2-dimensional dependence on $(\alpha_{1,t}^\phi, \alpha_{2,t}^\phi)$. Because of this, we can reduce \eqref{eq:JointDynamicsGeneral} to
\begin{equation}
\label{eq:JointDynamicsSpecific}
\begin{cases}
	d\alpha_{1,t}^\phi &= \frac{dt}{t} \left( \cos(\theta_t + \phi) - \alpha_{1,t}^\phi \right) \\
	d\alpha_{2,t}^\phi &= \frac{dt}{t} \left( \sin(\theta_t + \phi) - \alpha_{2,t}^\phi \right) \\
    d\theta_t &= \sigma dW_t - 2 c \left( \alpha_{2,t}^\phi \cos \theta_t - \alpha_{1,t}^\phi \sin \theta_t \right) \, .
\end{cases}
\end{equation}

Given the Markov process $(\theta_t, \alpha_{1,t}^\phi, \alpha_{2,t}^\phi)$, we can write the time-dependent Fokker--Planck equation for $$\mathbb{P}(\theta_t \in [\theta, \theta + d\theta), \alpha_{1,t}^\phi \in [\alpha_1^\phi, \alpha_1^\phi + d\alpha_1^\phi), \alpha_{2,t}^\phi \in [\alpha_2^\phi, \alpha_2^\phi + d\alpha_2^\phi)) \eqqcolon \nu_t(\theta, \alpha_1^\phi, \alpha_2^\phi) d\theta d\alpha_1^\phi d\alpha_2^\phi$$ as follows:
\begin{equation}
\label{eq:TimeDepFPEq}
\partial_t \nu_t(\theta, \alpha_1^\phi, \alpha_2^\phi) = - \nabla \cdot \left( \mathcal{F}_t(\theta, \alpha_1^\phi, \alpha_2^\phi) \nu_t(\theta, \alpha_1^\phi, \alpha_2^\phi) \right) + \frac{\sigma^2}{2} \partial_\theta^2 \nu_t(\theta, \alpha_1^\phi, \alpha_2^\phi) \, ,
\end{equation}
where $\nabla = \left( \partial_\theta, \partial_{\alpha_1^\phi}, \partial_{\alpha_2^\phi} \right)^\top$ and
\begin{equation}
\label{eq:DriftFP}
\mathcal{F}_t(\theta, \alpha_1^\phi, \alpha_2^\phi) = 
\begin{pmatrix}
- 2 c \left( \alpha_2^\phi \cos \theta  - \alpha_1^\phi \sin \theta  \right) \\
\frac{\cos(\theta + \phi)}{t} - \frac{\alpha_1^\phi}{t} \\
\frac{\sin(\theta + \phi)}{t} - \frac{\alpha_2^\phi}{t}
\end{pmatrix} \, .
\end{equation}

We now assume that \eqref{eq:TimeDepFPEq} reaches stationarity in the long-time limit, such that $\nu_t \rightarrow \nu_\infty$ and $\mathcal{F}_t \rightarrow \mathcal{F}_\infty$. In this limit, we observe that $\mathcal{F}_\infty$ effectively reduces to a one-dimensional vector field, and thus \eqref{eq:TimeDepFPEq} simplifies to
\begin{equation}
\label{eq:FPStationary}
0 = - \partial_\theta \left( \mathcal{F}_\infty(\theta, \alpha_1^\phi, \alpha_2^\phi) \nu_\infty(\theta, \alpha_1^\phi, \alpha_2^\phi) \right) + \frac{\sigma^2}{2} \partial_\theta^2 \nu_\infty(\theta, \alpha_1^\phi, \alpha_2^\phi) \, .
\end{equation}
As a consequence, $\alpha_1^\phi$ and $\alpha_2^\phi$ have effectively become parameters of the equation and characterise $\rho_{\text{inv}}$, as can be gathered from the form of the drift in \eqref{eq:AdditionSine}. Furthermore, in the long-time limit
\begin{equation}
\label{eq:ConnectionFP}
\nu_\infty(\theta, \alpha_1^\phi, \alpha_2^\phi) \equiv \rho_{\text{inv}}(\theta) \, ,
\end{equation}
as the empirical occupation measure converges in the long-time limit to the probability to find the self-interacting particle somewhere on the ring. Therefore, \eqref{eq:FPStationary} is equivalent to \eqref{eq:SIDStatFokkerPlanck}.
}

\section{Linear stability of $\rho_{\text{inv}} = (2 \pi)^{-1}$}
\label{appendix:linear}

We study the time-dependent Fokker--Planck equation
\begin{equation}
\label{eq:TimeDependentFokkerPlanckLinearStability}
\frac{\partial \rho_t(\theta)}{\partial t} = -\frac{\partial }{\partial \theta} \left( \rho_t(\theta) F_{\rho_t}(\theta) \right) + \frac{D}{2} \frac{\partial^2 \rho_t(\theta)}{\partial \theta^2} \, ,
\end{equation}
with 
\begin{equation}
\label{eq:PerturbationLinearStability}
\rho_t(\theta) = \frac{1}{2\pi} + \epsilon \eta_t(\theta) \, ,
\end{equation}
for $\epsilon \ll 1$ and $F_{\rho_t}$ defined in \eqref{eq:DriftRhoRing}. Notice that although \eqref{eq:TimeDependentFokkerPlanckLinearStability} is not valid in general as discussed in the main text, it is valid in the linear stability study that follows as we look at a perturbation in time only around the stationary solution $\rho_{\text{inv}} = (2\pi)^{-1}$ as in \eqref{eq:PerturbationLinearStability}.

Due to the normalisation of $\rho_t(\theta)$ it follows that $\eta_t$ has zero mean, i.e.,
\begin{equation}
\label{eq:PropertiesHigherMoodsLinearStability}
\int_0^{2 \pi} \eta_t(\theta) \, d\theta = 0 \, . 
\end{equation}
Additionally, $\eta_t$ is also $2\pi$-periodic, i.e., $\eta_t(0) = \eta_t(2\pi)$.

Replacing (\ref{eq:PerturbationLinearStability}) in (\ref{eq:TimeDependentFokkerPlanckLinearStability}) we get, at order $\epsilon$,
\begin{equation}
\label{eq:ProofStep1LinearStability}
\begin{split}
\frac{\partial \eta_t(\theta)}{\partial t} = - \frac{c}{\pi} \int_0^{2 \pi} \cos (\theta_1 - \theta + \phi) \eta_t(\theta_1) \, d\theta_1 + \frac{D}{2} \frac{\partial^2 \eta_t(\theta)}{\partial \theta^2} \, .
\end{split}
\end{equation}
We now expand the perturbation $\eta_t(\theta)$ in a Fourier series, i.e.,
\begin{equation}
\label{eq:FourierPerturbationLinearStabilityAnalysis}
\eta_t(\theta) = d(t) e^{i \theta} + d^*(t) e^{- i \theta} + \eta^{\perp}_t(\theta) \, , 
\end{equation}
where we only highlight the first oscillation mode.

We replace (\ref{eq:FourierPerturbationLinearStabilityAnalysis}) in (\ref{eq:ProofStep1LinearStability}) and obtain
\begin{equation}
\label{eq:ProofStep2LinearStabilityAnalysis}
\begin{split}
\left( \frac{\partial d(t)}{\partial t} e^{i \theta} + \frac{\partial d^*(t)}{\partial t} e^{-i \theta} + \frac{\partial \eta^{\perp}_t(\theta)}{\partial t}\right) &= - \frac{c}{\pi} \int_0^{2 \pi} \cos(\theta_1 - \theta + \phi) \left( d(t) e^{i \theta_1} + d^*(t) e^{-i \theta_1} + \eta_t^{\perp}(\theta_1) \right) \, d\theta_1 + \\
&\hspace{3cm} + \frac{D}{2} \left( - d(t) e^{i \theta} - d^*(t) e^{- i \theta} + \frac{\partial^2 \eta_t^{\perp}(\theta)}{\partial \theta^2} \right) \, .
\end{split}
\end{equation}
We only focus on the equation for the amplitude $d$ of the fundamental oscillatory mode, which reads
\begin{equation}
\label{eq:AmplitudeLinearStabilityAnalysis}
\begin{split}
\frac{\partial d(t)}{\partial t} e^{i \theta} &= - \frac{c d(t)}{\pi} \int_0^{2 \pi}\cos(\theta_1 - \theta + \phi) e^{i \theta_1} \, d\theta_1 - \frac{D d(t)}{2} e^{i \theta} \\
&= - \frac{c d(t)}{2 \pi} \int_0^{2 \pi} \left( e^{i \theta_1} e^{i(- \theta + \phi)} + e^{ - i \theta_1} e^{i(\theta - \phi)} \right) e^{i \theta_1} d \theta_1 - \frac{D d(t)}{2} e^{i \theta} \\
&= - c d(t) e^{i \theta} e^{- i \phi} - \frac{D d(t)}{2} e^{i \theta} \, .
\end{split} 
\end{equation}
Hence, the amplitude equation eventually becomes
\begin{equation}
\label{eq:AmplitudeFinalLinearStabilityAnalysis}
\frac{d d(t)}{d t} = - c d(t) e^{- i \phi} - \frac{D}{2} d(t) \, .
\end{equation}

It is interesting to notice that higher modes do not influence the linear stability. This can be seen extrapolating the equation for $\eta_t^{\perp}$ from (\ref{eq:ProofStep2LinearStabilityAnalysis}), which reads
\begin{equation}
\label{eq:HigherModesLinearStabilityAnalysis}
\begin{split}
\frac{\partial \eta_t^{\perp}(\theta)}{\partial t} &= -\frac{c}{\pi} \int_0^{2 \pi} \cos(\theta_1 - \theta + \phi) \eta_t^{\perp} (\theta_1) \, d\theta_1 + \frac{D}{2} \frac{\partial^2 \eta_t^{\perp}(\theta)}{\partial \theta^2} \\
&\stackrel{(\ref{eq:PropertiesHigherMoodsLinearStability})}{=} \frac{D}{2} \frac{\partial^2 \eta_t^{\perp}(\theta)}{\partial \theta^2} \, .
\end{split}
\end{equation}
If we now replace\footnote{$a_k = a_{-k}$ as $\eta_t^{\perp} \in \mathbb{R}$.}
\begin{equation}
\label{eq:FourierSeriesHighModesLinearStability}
\eta_t^{\perp}(\theta) = \sum_{k \geq 2} a_k(t) e^{i k \theta} \, ,
\end{equation}
into (\ref{eq:HigherModesLinearStabilityAnalysis}), we get
\begin{equation}
\frac{d a_k}{d t} = - \frac{D k^2}{2} a_k \, ,
\end{equation}
which has solution
\begin{equation}
a_k(t) = a_k(0) e^{- \frac{D k^2}{2}t} \, .
\end{equation}
Eventually, we find that 
\begin{equation}
\label{eq:BehaviourHighModesLinearStability}
\eta_t^{\perp}(\theta) = \sum_{k \geq 2} a_k(0) e^{- \frac{D k^2}{2} t} e^{i k \theta} \, ,
\end{equation}
which means that higher harmonics decay to zero exponentially fast in time for $D > 0$, and so everything is controlled by the first oscillation mode.
To study the spectrum of (\ref{eq:AmplitudeFinalLinearStabilityAnalysis}) we write $d(t) = b e^{\lambda t}$ and replace it in (\ref{eq:AmplitudeFinalLinearStabilityAnalysis}). Linear stability is guaranteed by $\operatorname{Re}(\lambda) < 0$, that is
\begin{equation}
\operatorname{Re}(\lambda) = \operatorname{Re}\left( - c e^{- i \phi} - \frac{D}{2} \right) < 0 \iff \frac{c}{D} \cos \phi \geq - \frac{1}{2} \, .
\end{equation}

\section{Connection to the mean-field XY model}
\label{appendix:XY}

We introduce the mean-field XY model (or $\equiv O(2)$ model) on a finite complete graph $\mathcal{G}$ of $N$ vertices. At each site $i$ of $\mathcal{G}$ there is a spin taking values in $[0,2\pi)$. The state space of the whole system is $[0,2\pi)^N$. The corresponding mean-field Hamiltonian $H_N: [0,2\pi)^N \rightarrow \mathbb{R}$ is
\begin{equation}
\label{eq:HamiltonianXYmodel}
H_N = - \frac{B}{N} \sum_{i,j \in \mathcal{G}} \cos (\theta^{j,N} - \theta^{i,N} + \phi) \, ,
\end{equation}
where $\theta^{i,N}$ represents the phase of spin $i$, and $\phi$ is a phase-shift affecting the two-body interaction between spins. In the following, we will set $B = - 2 c$, and $\phi = \pi$.

We now consider the dynamics of each site individually~\cite{DenHollander2000}, i.e.,
\begin{equation}
\label{eq:SDEInteractingSpins}
\begin{split}
d \theta_t^{i,N} &= - \frac{d H_N}{d \theta_t^{i,N}} + \sqrt{D} d W_t^{i,N} \\ 
&= \frac{2 c}{N} \sum_{j \in V} \frac{d}{d \theta_t^{i,N}} \cos (\theta_t^{j,N} - \theta_t^{i,N}) + \sqrt{D} d W_t^{i,N} \, ,
\end{split}
\end{equation}
where $\left( W_t^{i,N} \right)_{i=1}^N$ are $N$ independent Brownian motions on $S^1$ such that
\begin{equation}
\label{eq:CorrW}
\mathbb{E} \left[ dW_t^{i,N}, dW_{t+s}^{j,N} \right] = 2 \delta_{ij} \delta(s) \, .
\end{equation}

Following~\cite{Dean1996}, the Dean's equation, i.e., the evolution equation for the overall distribution $\rho_t^N$ of the spins given by
\begin{equation}
\label{eq:SDEInterSpinsDensity}
\rho_t^N(\theta) = \frac{1}{N} \sum_{i=1}^N \rho_t^{i,N} = \frac{1}{N} \sum_{i=1}^N \delta(\theta_t^{i,N} - \theta)  \, ,
\end{equation}
where $\rho_t^{i,N}$ is the density function of a single diffusing particle, reads
\begin{equation}
\label{eq:DeanEquationInterSpins}
\frac{\partial \rho_t^N(\theta)}{\partial t} = - \frac{\partial}{\partial \theta} \left( 2 c \rho_t^N(\theta) \int_0^{2 \pi} \sin (\theta_1 - \theta) \rho_t^N(\theta_1) \, d \theta_1 \right) + \frac{D}{2} \frac{\partial^2}{\partial \theta^2} \rho_t^N(\theta) - \frac{\partial}{\partial \theta} \left( \eta_t(\theta) \left( \rho_t^N \right)^{\frac{1}{2}} \right) \, ,
\end{equation}
where $\eta_t$ is an uncorrelated white noise such that
\begin{equation}
\label{eq:UncorrelatedWhiteNoise}
\mathbb{E} \left[ \eta_t(\theta), \eta_{t+s}(\theta') \right] = \frac{2}{N} \delta(\theta - \theta') \delta(s) \, .
\end{equation}

Remarkably, in the limit for $N \rightarrow \infty$ the multiplicative noise term in \eqref{eq:DeanEquationInterSpins} disappears and, along with $t \rightarrow \infty$, we recover the stationary Fokker--Planck equation \eqref{eq:SIDStatFokkerPlanck} for the SID. Hence, the infinite-time and infinite-size limit of a mean-field XY model is equivalent to the stationary limit of an SID.

\newtext{
\section{Details on Monte-Carlo simulations of SID in Eq.\ \eqref{eq:SIDEmp}}
\label{appendix:MonteCarlo}

We consider the simplest Euler--Maruyama discretised version of the SID in \eqref{eq:SIDEmp}, which is the discrete-time process $(\hat{\theta}_n)_{0 \leq n \leq N-1}$, with initial condition $\hat{\theta}_0 = 0$, defined by the recursive relation
\begin{equation}
\label{eq:EulerMaruyama}
\hat{\theta}_{n+1} = \hat{\theta}_n + \sqrt{\Delta t D} \, \xi_n + \Delta t \int_0^{2\pi} 2c \sin (\theta - \hat{\theta}_{n}) \hat{\rho}_n(\theta) \, d\theta \, ,
\end{equation}
where
\begin{equation}
\label{eq:EmpOccTimeDiscr}
\hat{\rho}_n (\theta) = \frac{1}{n} \sum_{\ell = 0}^{n-1} \delta({\hat{\theta}_\ell - \theta}) \, ,
\end{equation}
with $\hat{\rho}_0(\theta) = \delta(\theta)$, is the time-discretised version of the empirical occupation measure in \eqref{eq:EmpOcc}.
Having partioned the time interval $[0,t)$ in $N$ subintervals of length $\Delta t$, such that $0 = t_0 < t_1 < \cdots < t_{N-1} = t$, $\xi_n$ in \eqref{eq:EulerMaruyama} are defined as 
\begin{equation}
\label{eq:NoiseEulerMaruyama}
\xi_n = \frac{1}{\sqrt{\Delta t}} \left( W_{t_{n+1}} - W_{t_n} \right) \, ,
\end{equation}
i.e., independent and identically distributed Gaussian random variables with expected value $0$ and variance $1$. 

In practice, the space is also discretized. Thus, $\hat{\rho}_n$ in \eqref{eq:EmpOccTimeDiscr} is treated as a vector normalized to 1, and the integral in \eqref{eq:EulerMaruyama} is computed using a numerical scheme. In our simulations, we divide the interval $[0, 2\pi)$ into $100$ equally spaced subintervals and apply the trapezoidal rule to evaluate the integral.

We also note that it is computationally far more efficient to track $\hat{\rho}_n$ directly, rather than storing the entire trajectory and then performing the time integral as in, for example, \eqref{eq:SIDGeneral}.}

\bibliographystyle{ieeetr}
\bibliography{mybib}

\end{document}